\documentclass[iop,numberedappendix,appendixfloats]{emulateapj}
\newcommand{\sersic}{S\'{e}rsic }
\newcommand{\dev}{de Vaucouleurs }
\newcommand{\img}{${\it ^{^{0.5}}I814}$}
\newcommand{\imgiso}{${\it ^{0.5}I814_{26}}$}

\begin{document}

\title{The Inside-out Growth of the Most Massive Galaxies at $0.3<\lowercase{z}<0.9$}

\shorttitle{Inside-out Growth of BCGs}
\author{Lei Bai\altaffilmark{1},
H.~K.~C. Yee\altaffilmark{1},
Renbin Yan\altaffilmark{2},
Eve Lee\altaffilmark{3},
David G. Gilbank\altaffilmark{4},
E. Ellingson\altaffilmark{5},
L. F. Barrientos\altaffilmark{6},
M. D. Gladders\altaffilmark{7},
B. C. Hsieh\altaffilmark{8},
I. H. Li\altaffilmark{1} }
\email{leibai@gmail.com}
\altaffiltext{1}{Department of Astronomy \& Astrophysics, University of Toronto, 50 St. George St., Toronto, Ontario, M5S 3H4, Canada \email{leibai@gmail.com} }
\altaffiltext{2}{Department of Physics and Astronomy, University of Kentucky, 505 Rose Street, Lexington, KY 40506-0055, USA}
\altaffiltext{3}{Astronomy Department, University of California at Berkeley, Berkeley, CA 94720, USA}
\altaffiltext{4}{South African Astronomical Observatory, P.O. Box 9, Observatory, 7935, South Africa}
\altaffiltext{5}{Center for Astrophysics and Space Astronomy, Department of Astrophysical and Planetary Science, UCB-389, University of Colorado, Boulder, CO 80309, USA}
\altaffiltext{6}{Instituto de Astrof\'{\i}sica, Pontificia Universidad Cat\'olica de Chile, Avda. Vicu\~na Mackenna 4860, Macul, Santiago, Chile}
\altaffiltext{7}{Department of Astronomy and Astrophysics, University of Chicago, 5640 S. Ellis Ave., Chicago, IL 60637, USA}
\altaffiltext{8}{Institute of Astronomy and Astrophysics, Academia Sinica, P.O. Box 23-141, Taipei 106, Taiwan, Republic of China}

\begin{abstract}
We study the surface brightness profiles of a sample of brightest cluster galaxies (BCGs) with $0.3<z<0.9$.
The BCGs are selected from the first Red-sequence Cluster Survey and an X-ray cluster survey. 
The surface brightness profiles of the BCGs are measured using $HST$ ACS images, and the majority of them can be well modeled by a single \sersic profile with a typical \sersic index $n\sim6$ and a half-light radius $\sim$30 kpc. 
Although the single \sersic model fits the profiles well, we argue that the systematics in the sky background measurement and the coupling between the model parameters make the comparison of the best-fit model parameters ambiguous. 
Direct comparison of the BCG profiles, on the other hand, has revealed an inside-out growth for these most massive galaxies: as the mass of a BCG increases, the central mass density of the galaxy increases slowly ($\rho_{\rm1kpc} \propto M_{*}^{0.39}$), while the slope of the outer profile grows continuously shallower ($\alpha_{r^{1/4}} \propto M_{*}^{-2.5}$). 
Such a fashion of growth continues down to the less massive early-type galaxies (ETGs) as a smooth function of galaxy mass, without apparent distinction between BCGs and non-BCGs.  
For the very massive ETGs and BCGs, the slope of the Kormendy relation starts to trace the slope of the surface brightness profiles and becomes insensitive to subtle profile evolution.
These results are generally consistent with dry mergers being the major driver of the mass growth for BCGs and massive ETGs.
We also find strong correlations between the richness of clusters and the properties of BCGs: the more massive the clusters are, the more massive the BCGs ($M^{*}_{\rm bcg} \propto M_{\rm clusters}^{0.6}$) and the shallower their surface brightness profiles. 
After taking into account the bias in the cluster samples, we find the masses of the BCGs have grown by at least a factor of 1.5 from $z=0.5$ to $z=0$, in contrast to the previous findings of no evolution. 
Such an evolution validates the expectation from the $\Lambda$CDM model.

\end{abstract}


\section{Introduction}
The brightest cluster galaxies (BCGs) are the most massive galaxies in the Universe. 
They are the dominant galaxies that sit at the bottom of the potential well of galaxy clusters, the largest gravitationally bound structures. 
They typically are elliptical galaxies with most of their stellar population formed before $z>2$ \citep[e.g.,][]{Thomas05}.
Their extreme size, mass and unique environment provide strong constraints on galaxy formation and evolution models. 

In the standard picture of the Lambda Cold Dark Matter ($\Lambda$CDM) cosmology model, smaller galaxies are formed first from gas condensations in dark matter halos and then hierarchically merge with each other to form more massive ones. 
BCGs are therefore expected to form at a later time.
Their old stellar population, on the other hand, suggests the formation of the stars has long been halted in these systems.
This leads to the distinct evolution for the star formation history and the mass assembly history of BCGs:
the star formation was completed at higher redshift in individual galaxies which were then assembled together at later times through dry mergers, forming a massive galaxy.
Detailed galaxy evolution models in the cosmological context have succeeded in producing BCGs with surface brightness profile matching the observations \citep{Dubinski98} and have generally predicted a large mass growth with redshift, around a factor of three, since $z\sim1$ \citep{deLucia07}.

The observational evidence on BCG evolution, however, is still full of controversies.
The stellar mass evolution based on the direct photometric measurement of the BCGs has yielded conflicting results with little to no evolution \citep{Collins98,Whiley08,Collins09,Stott10} to strong evolution \citep{Aragon98}. 
The discrepancies likely arise from the bias in the cluster sample selection \citep{Burke00}.
A recent study by \citet{Lidman12} has found a factor of 1.8 mass growth, from $z=0.9$ to $z=0.2$, after taking into account this selection bias.

Another way to quantify the evolution of the BCGs is to study their profile and size evolution.
Because of their extended profiles, the photometric measurement of BCGs could miss a substantial fraction of the total stellar light, which results in uncertainties in their stellar mass evolution. 
The direct comparison of the profiles, could provide an independent and more complete examination of the BCG evolution.
The observational studies of this type, again, have seen conflicting results.
\citet{Bernardi09} found the size of the BCGs at $z\sim0.25$ to be 70\% smaller than their local counterparts.
\citet{Ascaso11} compared 20 BCGs at $0.3<z<0.6$ to a sample of local BCGs and found no evolution in their profile shape but a factor of two growth in their size.
On the other hand, \citet{Stott11} found little change either in the shape or in the size when comparing five high-$z$ BCGs at $0.8<z<1.3$ to BCGs at $z=0.25$.

Related to the size evolution of the BCGs, recent observations have found a strong size evolution for the massive early-type galaxies (ETGs) since $z\sim2$ \citep{Daddi05,vanDokkum08,Damjanov09,vanDokkum10,Patel13}.
The size evolution of these galaxies is much faster than their mass evolution. 
Since $z\sim2$, they typically have grown by a factor of $\sim4$ in size and by a factor of $\sim2$ in mass. 
The stronger evolution in size than in mass, $r\propto M^2$, has favored minor dry mergers as the physical driver of such growth, because major mergers will grow the size and mass at the same rate while minor mergers can grow the size much faster \citep[e.g.,][]{Naab09,Hilz13}.
Because the BCGs are the most massive ETGs, minor dry mergers should play an even bigger role in their mass growth than for the less massive ETGs.
This would seemingly imply that BCGs should also have a size growth twice as fast as their mass growth. 
So a mass growth of a factor of two from $z=0.5$ to $z=0$ predicted by the simulation of \citet{deLucia07} should result in a size growth of a factor of four. 
Such drastic size growth contradicts the small or even no growth in size found in the observations of BCGs.

The controversies revolving around the size evolution of BCGs reflect the difficulty in the accurate size measurement for these massive ETGs.
BCGs typically have a surface brightness profile that is more extended and shallower at larger radii than a \dev profile.
The measurement of the shallow outer region depends sensitively on the sky background level: small systematics in the measured sky value can cause a large change in the size measurement.
Even with the same data set, differences in the profile modeling and sky measurement can lead to discrepant results \citep[see more discussion in][]{Stott11}.
In addition to the compilations in the size measurements, there are also few high-redshift BCGs which have images deep enough for profile studies.
These issues have been the main obstacles in quantifying BCG evolution.

In this work, we have assembled one of the largest BCG samples at $0.3<z<0.9$, with {\it Hubble Space Telescope (HST)} images deep enough for the profile study beyond 50 kpc to tackle these problems.
We study the general profile properties of these BCGs with model fitting and investigate how the systematics affect the model-dependent measurements.
We also explore the option to directly compare the surface brightness profiles of galaxies.
Finally, we discuss our results in the context of formation of massive galaxies.
The structure of the paper is laid out as follows: in $\S 2$, we present the sample and the {\it HST} data, and in \S 3 we describe the model fitting of the surface brightness profiles.
In \S 4 we directly compare the profiles of galaxies and characterize their change with mass using model-independent parameters.  
We discuss the Kormendy relation of the BCGs and evolution of the BCG structural parameters in \S 5.  
Throughout the paper, when comparisons between galaxies of different redshifts are made, we always shift the compared properties,  e.g., surface brightness, magnitude, and stellar mass etc., to their corresponding value at redshift 0.5, taking into account both cosmic dimming and passive evolution.
These adjustments are estimated with the Spectral Energy Distribution (SED) model of luminous red galaxies given by \citet{Maraston09}.
A $\Lambda$CDM cosmology with $\Omega_{M} = 0.3$, $\Omega_{\Lambda} = 0.7$, and $H_{0} = 70$ km s$^{-1}$ Mpc$^{-1}$ is used throughout this work.

\section{DATA AND REDUCTION}
\subsection{BCG Sample}
In total, we assemble a BCG sample with 37 BCGs in the redshift range of $0.3<z<0.9$. 
These BCGs are from two different surveys, which are discussed in the following sections.
\subsubsection{RCS Cluster BCG Sample}
The main BCG sample used in this work is from 29 galaxy clusters selected from the first Red-Sequence Cluster Survey \citep[RCS1;][]{Gladders05}. 
The redshifts of the clusters, $z_{\rm rcs}$, are estimated from the color of their red sequence.
We select the clusters with redshifts $0.3<z_{\rm rcs}<0.9$.
All of these clusters are from the 48 RCS1 clusters which have been imaged with the {\it HST} Advanced Camera for Surveys ($HST$/ACS) in the Snapshot mode with the F814W filter (GO-10616, PI Loh).
Each image has an exposure time of 1440 s.

Of the 48 RCS1 clusters with {\it HST} images, 46 of them have redshifts in the range of $0.3<z_{\rm rcs}<0.9$.
We further exclude five clusters that have less than three red galaxy candidates within $30\arcsec$ from the cluster center.
These red galaxy candidates are defined as the galaxies with color within $\pm0.1$ mag of the predicted red sequence color at $z_{\rm rcs}$ and with a predicted stellar mass $> 10^{11} M_{\sun}$.
For the clusters with at least three red galaxy candidates, the BCG is selected as the brightest one.
These BCGs are all regular elliptical galaxies except one spiral galaxy and one elliptical with double nucleus.
We also exclude these two from our BCG sample.
After some trial profile fitting, we further exclude 10 more BCGs with bright neighboring sources that make the profile measurement of the BCG highly uncertain.
In the end, we have a sample of 29 BCGs from RCS1 clusters (Table 1).
The majority of these 29 clusters are among the richest clusters in the RCS cluster catalog, with a red-sequence richness parameter, $B_{\rm gc}$ \citep[see detail in][]{Yee99}, ranging between $\sim20 $-$ 2000$ $h^{-1.8}_{50}\mbox{Mpc}^{1.8}$. 
This translates into a cluster velocity dispersion range of $\sim 350$-$1200~\mbox{km~s}^{-1}$ or a cluster mass range of $0.7$-$20 \times 10^{14} M_{\sun}$ \citep{Yee03}.  

We assign the redshift of the BCG to be the same as the red-sequence redshift ($z_{\rm rcs}$) of its host cluster. 
Among the 29 BCGs, 5 of them have spectroscopic redshifts measured by the Inamori Magellan Areal Camera and Spectrograph (IMACS) on the Baade 6.5 m Magellan telescope as part of an extensive spectroscopic survey for a subset of RCS1 clusters.
The full details of this spectroscopic survey will be reported by R. Yan et al. (2014, in preparation). 
In addition to these five BCGs, three more BCGs have spectroscopic redshifts from the Sloan Digital Sky Survey (SDSS).
For these eight BCGs, their spectroscopic redshifts ($z_{\rm spec}$) agree very well with $z_{\rm rcs}$, with a scatter of $\sim0.018$ and the largest difference of $0.034$. 
This corresponds to a $\sim4\%$ accuracy for $z_{\rm rcs}$. 
This number is smaller than the $\sim10\%$ accuracy for $z_{\rm rcs}$ reported by \citet{Gilbank07} for clusters with $z>0.7$.
The better accuracy for $z_{\rm rcs}$ of our cluster sample is mostly due to the lower-redshift range of our cluster sample, which means better photometry and better $z_{\rm rcs}$. 
In the rest of the paper, we use $z_{\rm spec}$ as the redshift of the BCG whenever it is available and $z_{\rm rcs}$ otherwise.
We note that the main results of this paper are not affected by the small uncertainties in $z_{\rm rcs}$.

\subsubsection{X-Ray Cluster BCG Sample}
In addition to the RCS cluster BCG sample, we also include eight more BCGs with $0.3<z<0.6$ from another ACS Snapshot survey with F814W filter (GO-10152, PI Donahue).
This survey targeted 25 clusters randomly drawn from a flux-limited X-ray cluster sample \citep{Mullis03}.
The sample's detailed properties were reported in \citet{Hoekstra11}.
The exposure times of these images are typically $\sim2000$ s, slightly deeper than the {\it HST} image of the RCS sample.

Although the redshifts of these clusters have been spectroscopically confirmed by \citet{Mullis03}, the spectroscopic data were only listed with respect to the central coordinates of the X-ray emission, not with the individual galaxies.   
Because of the large uncertainties in the X-ray centroid position (typically $10\arcsec$-$20\arcsec$) and the fact that X-ray emission is not always centered at the BCGs, the identification of the BCGs becomes ambiguous in many clusters. 
To avoid mistaking foreground galaxies as higher-redshift BCGs, we only selected BCGs that satisfy the following conditions: (1) they are elliptical galaxies; (2) they have either SDSS coverage or two-band {\it HST} photometry which we can use to confirm the BCGs as the brightest galaxies on the red sequence; (3) they are located within 250 kpc of the X-ray centroids.
We also exclude a few BCGs with neighboring galaxies of comparable brightness because the profile-fitting results of these systems are less reliable. 
In total, there are eight clusters with unambiguous and dominant BCGs (Table 2).
In all cases the BCGs we identified are consistent with the BCG identified by \citet{Hoekstra11}. 

\subsection{Early-Type Galaxy Sample}
For comparison, we also assemble a sample of ETGs in the RCS clusters with the {\it HST} images and analyze them in the same way as we do for the BCGs.
For ETGs, we only select the ellipticals with spectroscopic redshifts and with color matching the red-sequence color of the cluster. 
After profile fitting, we further discard any galaxies with a \sersic index less than 3 to ensure they are ellipticals.
In the end, we have 34 ETGs in the sample. 
This sample size is comparable to the size of our BCG sample.

\begin{deluxetable}{lrrrrr}
\tablewidth{0pt}
\tablecolumns{6}
\tabletypesize{\scriptsize}
\tablecaption{BCG Sample from the RCS Cluster Survey}
\tablehead{
Cluster & R.A. of BCG & Decl. of BCG & $z_{\rm spec}$ & $z_{\rm rcs}$ & $B_{\rm gc}\footnotemark[1]$
}
\tablecomments{Column 1: galaxy cluster; Columns 2 and 3: R.A. and decl. (J2000) of the BCGs; Column 4: spectroscopic redshift of BCGs; Column 5: photometric redshift of the cluster from the RCS catalog; Column 6: richness paramater $B_{\rm gc}$.} 
\startdata
\hline
RCS$1102-05$&11:02:59.2&$-05$:21:09.2&0.321&0.333& 620$\pm$192\\
RCS$2239-60$&22:39:54.8&$-60$:44:46.7&-&0.331&1006$\pm$229\\
RCS$0444-28$&04:44:08.7&$-28$:20:16.7&0.338&0.355&1002$\pm$225\\
RCS$0351-09$&03:51:39.9&$-09$:56:26.5&-&0.351& 742$\pm$204\\
RCS$0518-43$&05:18:33.8&$-43$:25:10.7&-&0.352& 699$\pm$181\\
RCS$1102-03$&11:02:33.0&$-03$:19:04.8&-&0.362&1194$\pm$245\\
RCS$0224-02$&02:24:03.4&$-02$:28:16.0&-&0.380& 661$\pm$196\\
RCS$0515-43$&05:15:37.2&$-43$:25:14.1&-&0.387& 950$\pm$225\\
RCS$0928+36$&09:28:21.2&$+36$:46:28.4&0.393\footnotemark[2]&0.407&1344$\pm$257\\
RCS$1452+08$&14:52:27.1&$+08$:34:54.7&0.396\footnotemark[2]&0.430& 620$\pm$192\\
RCS$1319-02$&13:19:12.2&$-02$:07:11.1&-&0.432&  20$\pm$114\\
RCS$1323+30$&13:23:34.1&$+30$:22:49.2&0.462\footnotemark[2]&0.435&2063$\pm$308\\
RCS$0511-42$&05:11:29.0&$-42$:35:12.3&-&0.473& 752$\pm$185\\
RCS$0518-43$&05:18:55.6&$-43$:15:06.1&-&0.509& 682$\pm$180\\
RCS$1107-05$&11:07:54.1&$-05$:16:39.5&-&0.531& 557$\pm$177\\
RCS$0519-42$&05:19:19.6&$-42$:47:51.7&-&0.557& 432$\pm$156\\
RCS$2316-00$&23:16:55.3&$-00$:11:47.8&-&0.561& 688$\pm$179\\
RCS$0350-08$&03:50:27.2&$-08$:54:56.3&-&0.566&1112$\pm$267\\
RCS$1108-04$&11:08:14.7&$-04$:30:50.2&-&0.586& 545$\pm$174\\
RCS$1446+08$&14:46:54.7&$+08$:27:04.7&0.630&0.646& 701$\pm$178\\
RCS$1419+53$&14:19:12.1&$+53$:26:11.9&-&0.634&1174$\pm$216\\
RCS$1104-04$&11:04:40.1&$-04$:44:58.5&0.641&0.654&1140$\pm$256\\
RCS$2342-35$&23:42:19.3&$-35$:34:15.5&-&0.680& 692$\pm$220\\
RCS$1107-05$&11:07:24.1&$-05$:23:19.7&-&0.715&1009$\pm$283\\
RCS$2152-06$&21:52:48.4&$-06$:09:36.3&-&0.731& 934$\pm$241\\
RCS$1450+08$&14:50:40.6&$+08$:40:43.4&-&0.786&1026$\pm$207\\
RCS$1122+24$&11:22:25.8&$+24$:22:29.9&-&0.813&1200$\pm$284\\
RCS$1620+29$&16:20:10.1&$+29$:29:24.1&-&0.828& 806$\pm$211\\
RCS$0519-44$&05:19:40.3&$-44$:02:21.1&0.831&0.828& 531$\pm$249\\
\enddata
\label{tab_rcs}
\footnotetext[1]{$B_{\rm gc}$ is in unit of $h^{-1.8}_{50}\mbox{Mpc}^{1.8}$.}
\footnotetext[2]{Measurements from SDSS.} 
\end{deluxetable}

\begin{deluxetable}{lrrr}
\tabletypesize{\scriptsize}
\tablecolumns{4}
\tablecaption{BCG Sample from the X-Ray Cluster Survey}
\tablehead{
Cluster & R.A. of BCG & Decl. of BCG & $z_{\rm spec}$
}
\tablecomments{Column 1: galaxy cluster; Columns 2 and 3: R.A. and decl. (J2000) of the BCGs; Column 3: spectroscopic redshift of BCGs.}  
\startdata
\hline
RXJ$0110+19$&01:10:18.2&$+19$:38:19.1&0.317\\
RXJ$0841+64$&08:41:07.7&$+64$:22:26.2&0.342\\
RXJ$1540+14$&15:40:53.9&$+14$:45:56.4&0.441\\
RXJ$0926+12$&09:26:36.7&$+12$:43:04.2&0.489\\
RXJ$2328+14$&23:28:52.3&$+14$:52:43.4&0.497\\
RXJ$0056-27$&00:56:57.0&$-27$:40:29.7&0.560\\
RXJ$0847+34$&08:47:11.7&$+34$:48:52.5&0.560\\
RXJ$1354-02$&13:54:17.2&$-02$:21:59.0&0.566\\
\enddata
\label{tab_xray}
\end{deluxetable}

\subsection{ACS Data}
All the data used in this work are taken with the Wide Field Channel (WFC) of ACS with the F814W filter.
The images were retrieved from the Hubble Legacy Archive (HLA) and processed with the MultiDrizzle pipeline with standard parameters. 
Multiple exposures were used to remove cosmic rays and bad pixels, and corrected for the geometric distortion and offsets before being coadded together.  
The final image products have a pixel scale of $0\farcs05$, which preserves the noise properties of the original images and allows for an accurate profile fitting.
The spatial resolution of the images is $\sim 0\farcs15$.

The WFC array is composed of two $2048\times4096$ CCDs with a full size of the field of view about $3\arcmin\times3\arcmin$.
BCGs are typically placed off center to avoid CCD gaps.
Because all the data used in this study were taken before the Servicing Mission 4 of the {\it HST}, there is a quadrant-to-quadrant bias jump that cannot be removed by the standard calibration. 
This causes discontinuity in sky background across the quadrants.
For the BCGs which sit across different quadrants, we first measure the sky background level of each individual quadrant and then adjust the sky level accordingly to correct for the bias jump.
In theory, the size of the CCD suggests we should be able to measure the sky background more than 500 kpc away from the BCGs at $z=0.5$.
In practice, however, due to the uncertainty in the flatness of the sky background and the fact that BCGs are off center in the images, we typically can only measure the sky background reliably around $\sim$ 300 kpc from the BCGs.

\section{MODEL FITTING OF SURFACE BRIGHTNESS PROFILES}
\subsection{Image Preparation for Profile Fitting}
To fully utilize the two-dimensional information of the galaxy surface brightness profile, we use GALFIT \citep{Peng10} to model our BCGs.  
Each BCG is cut into a $40\arcsec\times40\arcsec$ stamp image, and we run GALFIT using this stamp image.
This is about 240 kpc in size for BCGs with the median redshift of our sample.
In a few cases with large BCGs, we use $60\arcsec\times60\arcsec$ stamp images to accommodate their large size. 
The large fitting size we use here is essential to ensure the proper fitting of BCGs that could have a half-light radius $\sim100$ kpc. 
On the other hand, images bigger than $60\arcsec\times60\arcsec$ start to see considerable deviation of the sky background from a flat constant and result in increased uncertainties in the profile fitting.

For each stamp image, we visually inspect the image and manually mask out all the nearby sources. 
Typically, $\sim25\%$ of the pixels are masked out.
In a few cases where the BCGs overlap with other galaxies with comparable brightness, we also test the option of leaving the close neighbor unmasked and running GALFIT on both galaxies simultaneously. 
The results are not significantly different from modeling the BCG alone.

GALFIT convolves a model image with the point spread function (PSF) before comparing it with the data. 
A good PSF can be constructed with a well-isolated star in the image that is bright enough to constrain the PSF well but not too bright that it is saturated.
Because there is not always such a star in the field close to BCGs, we use the PSF model produced by Tinytim \citep{Krist11}. 
We generate an instrument-dependent PSF at the CCD position of each BCG using Tinytim, but we do not apply geometric distortion to the PSF because MultiDrizzle already corrects for distortion. 
Because the theoretical PSF is always sharper than the actual observed one, we further convolve the PSF with a Gaussian function to match the FWHM of the model PSF to the actual stars in the images. 
In the fields for which we do have a good star to use as a PSF reference, we get consistent results using either modeled PSF or the observed one.
To be consistent, our final results are all based on modeled PSFs.

The determination of the sky background level is crucial for proper morphological fitting with GALFIT.
When the galaxy has very extended emission, as BCGs normally do, the size measured by GALFIT is quite sensitive to the sky value.
To minimize the uncertainty caused by the sky background value, we manually select regions around each BCG that are free of sources and are at least 300 kpc away from BCGs. 
We calculate the sky background in those regions by performing iterative 3-$\sigma$ clipped mean and input this value to GALFIT as a fixed parameter.
Because the fitting region we select is quite large, for most BCGs the fitting results obtained using fixed sky background values are consistent with the results obtained by leaving it as a free parameter.
However, in a few cases, when a large number of pixels are masked out due to a crowded field, the manually measured sky background gives a more reliable fit.

\subsection{Model Selection}
Because the morphological analysis is conducted using model-dependent parameters, the choice of analytic models to fit the galaxy profile is very important.  
Many previous studies have shown that typical BCGs, unlike less massive ETGs, have surface brightness profiles deviating substantially from the \dev profile.
They tend to have more extended low surface-brightness envelopes and more peaked central profile, often better described by a \sersic profile with a \sersic index greater than 4 (the \dev profile is a special case of the \sersic profile with a \sersic index of 4), or a combination of two profile components.
In this work, we test three different models for the BCG brightness profile:
a single \sersic profile and two two-component models, one with two \dev profiles and one with \dev $+$ exponential profile (2deV and deV+Exp, henceforth).
Some studies have interpreted deV+Exp model as bulge+disk model.
This is because a face-on classical disk also has an exponential profile, a special case of the \sersic function with $n=1$. 
In our case, the exponential profile is only used to model a shallow light profile and should not be considered as evidence of ``disk" in these galaxies. 
We also have experimented with a \sersic+Exp model. 
This is the model used by \citet{Ascaso11}. 
However, we conclude that this model has too much freedom in fitting, and the results are not very robust compared to all other models; so we do not include it in this paper.  

A single \sersic profile model has seven free parameters: the $x$ and $y$ coordinates of the centroid, the \sersic index $n$, the effective radius $r_{e}$ that encloses half of the total flux, the normalization of the profile, the axis ratio $b/a$, and the position angle of the ellipse.  
Therefore, each component of the two-component models has 6 free parameters and altogether 12 free parameters for each model. 
The significance of the two-component models is not just the more fitting freedom of the radial profile but also the ability to decouple azimuthally distinct components, e.g., the inner and outer components, which may have different ellipticities and position angles \citep{Gonzalez05}. 
Although a single \dev profile does not reproduce the BCG's profile very well, we still include it in our analysis, mostly for comparison purposes. 

\begin{figure*}
\epsscale{1.2}
\plotone{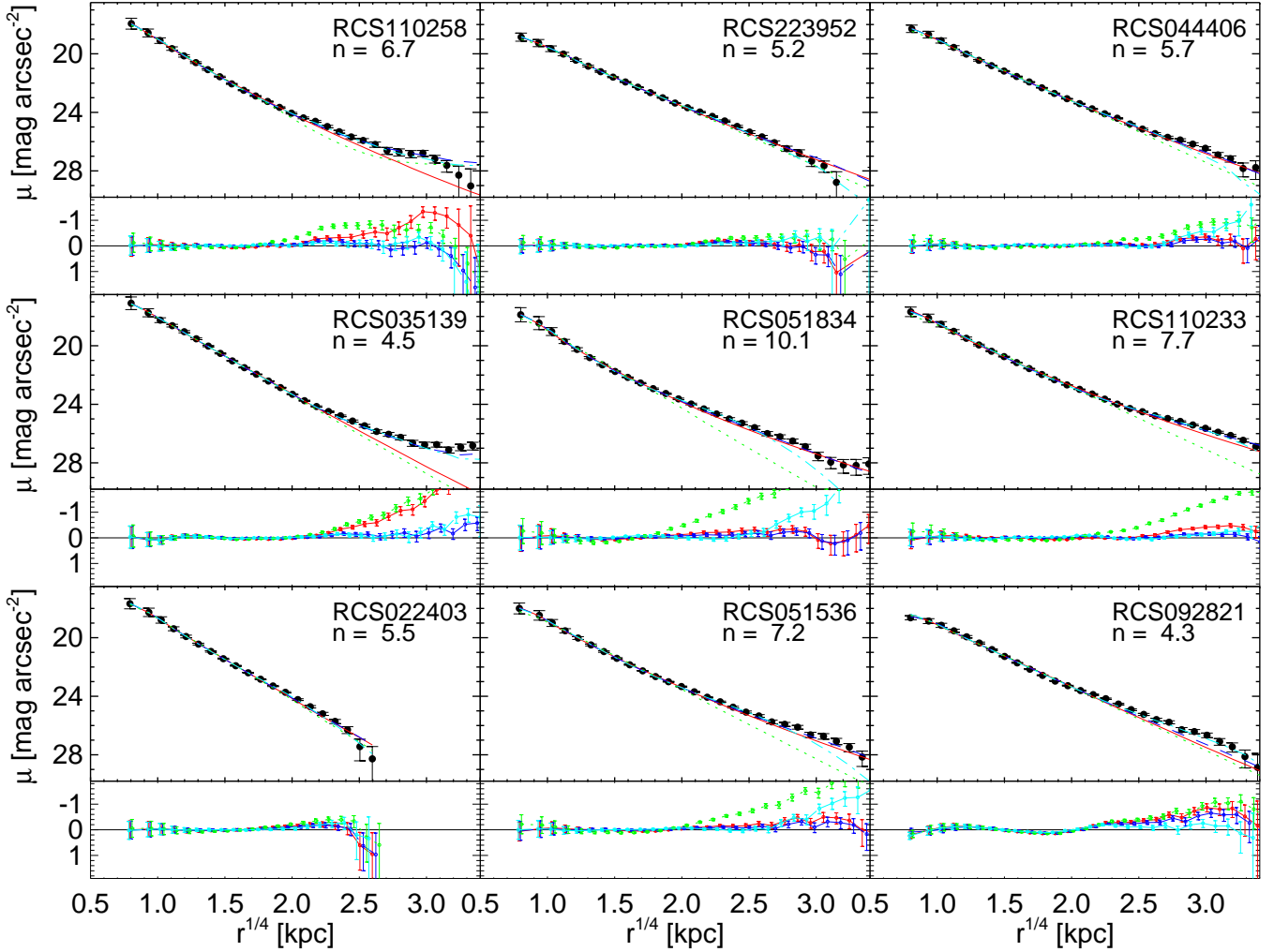}
\caption{The 1D surface brightness profiles of the BCGs and the best-fit models.
Only a subset of the BCG sample is shown here. 
The profiles for the rest of the sample are shown in Appendix B. 
Black filled circles are the data.
The red solid curve is the single \sersic model, the blue dashed curve is the 2deV model, the cyan dash-dotted curve is the deV+Exp model, and the green dotted curve is the single deV model.
The residuals of the fittings for each model are shown in the bottom panels.
}
\label{fig_profile}
\end{figure*}

\subsection{Fitting Results}
In almost all cases both single \sersic and 2deV models give robust and sensible fits without any fine tuning.
The deV+Exp model is less robust, and in many cases, we have to fix the centroid of the exponential profile to be the same as the \dev profile in order to have sensible results.
The best-fit parameters of each model can be found in Appendix A. 
Almost all of the model fits give a reduced $\chi^2$ very close to 1.
However, a unity $\chi^2$ value given by GALFIT does not necessarily mean a good fit.
This is because the non-Gaussianity of the noise in a real astronomical image invalidates the rigorous meaning of the $\chi^2$ \citep{Peng10}, and the absolute value of the reduced $\chi^2$ is not a good measure of the goodness of the fits. 
By the same token, small differences in the reduced $\chi^2$ cannot be reliably used to evaluate which model choice is better.

A more intuitive way to check the goodness of fit is to compare the one-dimensional (1D) profile of the real data to that predicted by the best-fit models. 
In Figure~\ref{fig_profile}, we show the 1D surface brightness profiles of a subsample of BCGs along with the best-fit models.
The same plot for the rest of the BCGs in our sample is included in Appendix B.
The profiles are extracted from the ellipses defined by the best-fit single \sersic parameters and are plotted as a function of $r^{1/4}$, where $r$ is measured along the major axis of the ellipse.
A \dev profile will show as a straight line in this plot except for the central region where the convolution of the PSF flattens the profile.
We extract the profiles from simulated galaxies of different models provided by GALFIT using the same ellipse defined by the best-fit \sersic model.
These model profiles are overplotted as colored curves on top of the data.
The residuals of the fits are plotted in the bottom panels of the figure.
As shown in the plot, a single \sersic model generally provides good fit for most of the BCGs, all the way from the central part to the outer region. 
In contrast, a single \dev profile often fails to reproduce the extended outer envelopes and underestimates the central brightness. 
The 2deV model generally agrees very well with the single \sersic model and in some cases shows a slight advantage in fitting very extended low-brightness outer region. 
This reflects the advantage of using the extra degrees of freedom the two-component models allow in fitting both inner and outer profiles simultaneously. 
The 1D profiles also suggest that the deV+Exp fits are slightly inferior to the 2deV fits and they are less robust.
In a few BCGs, all the models fail to reproduce profiles at large radii.
This is due to the residuals from the nearby bright neighbors even after masking or simultaneous profile fitting. 

The total magnitudes given by the best-fit \sersic and 2deV models are almost identical, with an average difference of $0.03\pm0.16$ mag.
The total magnitudes given by deV+Exp models, on the other hand, show a bigger discrepancy from the single \sersic and 2deV results, with an average difference of $0.16\pm0.25$ mag.
We note that these model total magnitudes are all much brighter, by almost 1 mag, than magnitudes given by general purpose photometric software, e.g., the mag\_best given by SExtractor.
This strongly suggests the general purpose photometric measurement can seriously underestimate the magnitude of galaxies with more extended profiles like BCGs.

In some cases where the galaxy has extra light at large radii, a two-component model fitting, especially the 2deV model, does give slightly better results than a single \sersic model.
However, at these radii the uncertainties in our sky background estimate start to affect the profile considerably, making the measurements of the extended envelopes dubious. 
If the extended envelope does exist, it probably is more closely related to the intracluster light (ICL) of stars that have been stripped off from the infalling galaxies \citep{Gonzalez05} rather than the galaxies themselves. 
Because of these uncertainties and given the fact that we do not see compelling evidence for the superiority of the two-component models compared to the single \sersic model in our data, we will focus our discussion on the single \sersic fitting results in the rest of the paper.

\begin{figure}
\epsscale{1.2}
\plotone{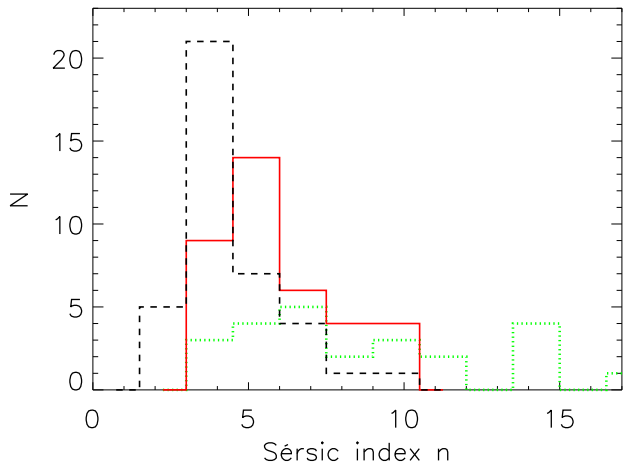}
\caption{Histogram of the \sersic index of the BCGs fitted with the single \sersic model (the red solid histogram).
The \dev profile has a \sersic index of 4.
The BCGs all have an index greater than 3, with a median value of 5.7.
The ETGs, shown as the black dashed histogram, have a median index of 3.9. 
After excluding ETGs with $n<3$, the median value increases to 4.3, but it is still smaller than the median value of the BCGs.
The green dotted histogram is the local BCG sample ($z\sim0.1$) from \citet{Gonzalez05}.
The indices of the local BCGs have higher median value (7.5) than the BCGs of this work, with $0.3<z<0.9$.
}
\label{fig_ndist}
\end{figure}

In Figure~\ref{fig_ndist}, we plot the histogram of the \sersic index for the single \sersic model of our BCGs sample.
All of them have an index value greater than 3, and the median of the sample is 5.7. 
As a comparison, the ETGs have a median index value of 3.9. 
If we exclude ETGs with $n<3$ to ensure they are truly ETGs, the median value increases to 4.3, but it is still smaller than that of the BCGs. 
The size of our BCGs, which is measured by the half-light radius $r_{e}$ of the single \sersic model along the major axis, ranges from $\sim$10-150 kpc.
The median size of the BCGs is $\sim30$ kpc, much larger than the median size of the ETGs, which is about 3 kpc. 

\subsection{Issues with Model Parameter Comparisons}\label{sec_issue}
Since a single \sersic model can fit the BCG profiles very well, it is sensible to study the properties of galaxies using only the best-fit model parameters.
This has become common practice for galaxy profile studies and is based on the belief that model parameters can fully characterize the profiles when the model itself has been shown to be a good fit of the observed profile.
For example, we can compare directly the model parameters of our BCG sample to those of the local BCG given by \citet{Gonzalez05}.
From $z=0.5$ to $z=0$, the median \sersic index, $n$, of BCGs has increased from 5.7 to 7.5 (see Figure~\ref{fig_ndist}), and the median half-light radius, $r_{e}$, increased from $\sim30$ to $\sim200$ kpc.
There are only two BCGs in our sample with $r_{e} \sim 100$ kpc, while more than half of local BCGs have $r_{e} > 100$ kpc.
Such a big difference in the half-light radius of the BCG samples seems to suggest a strong evolution in the BCG size from $z=0.5$ to $z=0$.
However, there are several issues related to these types of comparisons.

The first issue is the systematic uncertainties in the profile measurements, in particular, the uncertainties in the sky background measurement from different studies.
Both \citet{Gonzalez05} and \citet{Brown97} utilized special observational techniques in order to achieve accurate sky-level measurements out to large radii $\ge 600$ kpc.
This is very important to recover faint extended emission from BCGs and/or ICL.   
Most of the BCGs in our sample only have reliable sky background measurements within a 300 kpc radius, which is limited by the flatness of the sky background and the size of the {\it HST} camera field-of-view.
In theory, the actual size of the galaxy we can measure with GALFIT is not limited by the size of the image.
GALFIT can fit the sky background and the galaxy profile simultaneously and recover the real sky background level by extrapolating the best fit model.
However, in practice, uncertainties in the sky measurement in a limited-size image can result from the deviation of the sky from the assumed model, which in our case is assumed to be flat (higher order of curvature in the sky model can cause large uncertainties), and this can directly affect the ability of GALFIT to measure the sky accurately and cause biases in the size measurement of galaxies.

To test this possibility, we use GALFIT to generate two mock galaxies with \sersic surface brightness profiles.
The \sersic parameters of these two mock galaxies are the same as the two local BCGs, A2969 and A2730, from \citet{Gonzalez05}, after they have been moved from $z=0.1$ to the median redshift of our BCG sample, $z=0.5$, taking into account both cosmic dimming and passive evolution with SED models from \citet{Maraston09}.
These two BCGs are chosen because A2969 has \sersic parameters similar to the median values of the local BCG sample: $r_{e}= 222$ kpc, $n=7.3$, and the magnitude in $I813$ band at $z=0.5$ (\img) = 16.93 mag and A2730 is more on the extreme side of large size, high \sersic index, and high mass: $r_{e}$= 1364 kpc, $n$=14.4, and \img = 15.73 mag.  
We add Poisson noise to the mock galaxies and insert them into one of our blank sky ACS images. 
We then process these images with GALFIT using the same procedure we used for our BCGs and measure the sky background about 300 kpc from the mock galaxy to feed into GALFIT.
The best-fit parameters given by GALFIT are $r_{e}$= 166 kpc, $n$=6.9 for the mock A2969 BCG and $r_{e}$= 782 kpc, $n$=13.3 for A2730. 
In both cases, the measured $r_{e}$ is much smaller than the input value, and the larger the size, the greater the reduction (27\% and 43\% loss for the small and big galaxy, respectively).
The measured \sersic index values are smaller as well. 
This implies that the systematics in sky measurement do contribute partly to the difference we see in the size of the local and our BCGs, but they are not large enough to explain the entire difference we observe: even with a reduction of $\sim30\%$, the median size of the local BCGs is still much larger (at least five times more) than the median size of 30 kpc we measured in our BCG sample.
However, we need to caution that the mock galaxies were generated using \sersic profiles, and this may help GALFIT to better recover the input parameters. 
It is not clear how well GALFIT will do when there is departure of the real galaxy profiles from the assumed fitting model.
The systematic error measured by this method, therefore, will probably only account for part of the real systematics. 
These uncertainties hinder us from quantifying the real evolution in the BCG profiles. 

The second issue is the covariance between the fitting parameters, $r_{e}$ and $n$.
For a \sersic profile, its shape depends on both its $r_{e}$ and $n$.
However, when fitting a galaxy, many combinations of $r_{e}$ and $n$ can give reasonably good fits, and the two parameters couple approximately as $n\propto \log r_{e}$ \citep[e.g.,][]{Graham97}.
A small change in the best-fit $n$ can therefore produce a big change in $r_{e}$. 
The uncertainties in these parameters are dominated by the covariance and are larger than the typical uncertainties given by the fitting program. 
This makes it difficult to compare $r_{e}$ between different studies.
Even within the same study, this covariance makes it harder to interpret the significance of the difference in $r_{e}$ of different galaxies.
  
\section{ANALYSIS OF SURFACE BRIGHTNESS PROFILES}
\subsection{Direct Comparison of Surface Brightness Profiles}
\begin{figure}
\epsscale{1.2}
\plotone{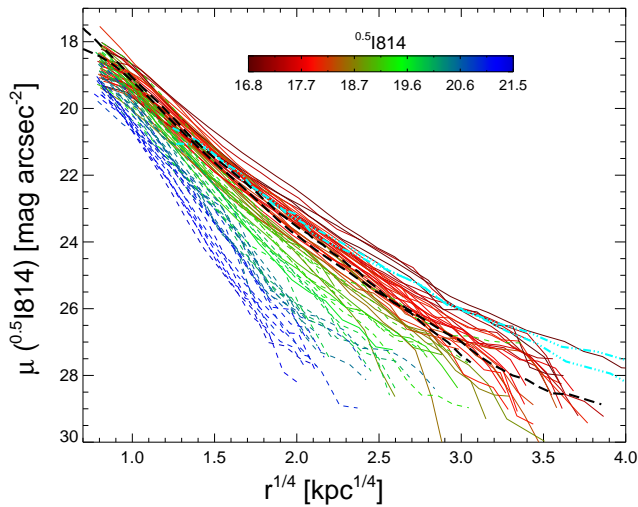}
\caption{Surface brightness profiles of BCGs and ETGs as a function of $r^{1/4}$.
The solid profiles are BCGs, and the short dashed ones are ETGs.
The profiles are color-coded by the integrated \img~ magnitudes of the best-fit \sersic model, as shown by the color bar.
The cyan dash-dot-dot-dot curves are the profiles of local BCG A2629 and A2730 from \citet{Gonzalez05}.
The black long dashed curves are M87 and M49, the two brightest ETGs of the Virgo cluster, from \citet{Kormendy09}.
All the surface brightness values have been corrected to the value at $z=0.5$ by taking into account cosmological dimming and passive evolution.
}
\label{fig_stack_profile_r25}
\end{figure}

In addition to the problems related to the \sersic model parameter comparison, the interpretation of the change in these parameters can also be problematic and sometimes misleading.
An increase in the half-light radius, $r_{e}$, can be due to the galaxy growth on all scales and/or a less concentrated light distribution; an increase in the \sersic index, $n$, at the same time, would imply a more peaked distribution in the central region and more extended distribution in the outer region.
Given the simultaneous increases in both $r_{e}$ and $n$, it is not clear how the surface brightness profiles of BCG change.  
A more straightforward way to investigate the profile properties of a population is to directly compare their profiles.  
In Figure~\ref{fig_stack_profile_r25}, we show the surface brightness profiles of all the BCGs in our sample as a function of $r^{1/4}$.
The surface brightness profiles are calculated directly from the image, assuming the centroid, ellipticity and position angle given by the best-fit \sersic model. 
This is the only model-dependent information we used in constructing the profiles. 
We then shift all the profiles to redshift 0.5 for direct comparison, taking into account both cosmic dimming and passive evolution of galaxies.
All the profiles are color-coded by the integrated \img~ magnitudes from the best-fit \sersic model.

Although the BCGs in our sample span about 2 mag in the total integrated \img, their surface brightness profiles are very similar to each other and form a relatively tight bundle in the plot.  
The more massive BCGs have slightly shallower slope but the difference is small, especially for BCGs with \img$<$ 18.5 mag ($\sim10^{12}M_{\sun}$).
The scatter in the surface brightness at fixed radius among different profiles is relatively small across a large range of radii.
It is about 0.41 mag arcsec$^{-2}$ from the centers to the inner 10 kpc and increases slightly to 0.66 mag arcsec$^{-2}$ at 40 kpc.
The ETGs in our sample, on the other hand, show much steeper decrease of brightness compared to BCGs and the slopes show stronger dependence on the total magnitude of galaxies.
However, the central surface brightness of ETGs is only slightly fainter than that of the BCGs.
This strongly suggests that the difference in the light distribution of BCGs and ETGs is mostly seen at larger radii, not in the central surface brightness.

Furthermore, for a couple of BCGs with \img $> 18.5$, their profiles are indistinguishable from the ETGs with similar total magnitudes.
This, plus the fact that BCG and ETG profiles form a continuous sequence, seems to favor the difference in profiles being primarily due to the difference in the stellar masses of the galaxies. 
Whether the galaxy is a BCG, the most dominant galaxy in the dark matter halo, may not be relevant.
However, this is not to say that the dark matter halo plays no role in forming the brightness profile of these massive galaxies.
Some, if not all, non-BCG ETGs were likely once the dominant galaxies in their local dark matter halo before they merged with more massive ones.
Thus, their evolution may not be fundamentally different from BCGs.

How do our BCGs' surface brightness profiles compare directly to the local BCGs?
In Figure~\ref{fig_stack_profile_r25} we plot the profiles of two local BCGs along with our BCGs.
Two local BCGs are of A2969 and A2730 from \citet{Gonzalez05}.
As mentioned in \S \ref{sec_issue}, A2969 BCG represents the median local BCG, and A2730 BCG is more on the extreme side. 
The half-light radii of these two BCGs are about 7-40 times bigger than the typical high-$z$ BCGs in our sample.
Despite their much larger sizes, their profiles follow the BCG profiles in our sample out to $\sim$ 20 kpc and become moderately flatter in outer regions.       
Only when reaching beyond 100 kpc do their profiles start to significantly depart from high-$z$ BCG profiles, showing slow flattening of the profile.
Between these two local BCGs, although their sizes differ by a factor of five, their profiles are very similar to each other out to $\sim$ 150 kpc.
This indicates the much bigger half-light radius of the A2730 BCG is mostly driven by its extended outer envelope.

As another independent check, we also compare the high-$z$ BCGs to the two brightest ellipticals, M87 and M49, from the Virgo cluster, taken from \citet{Kormendy09}.
The Virgo cluster, unlike A2969 and A2730, which are rich clusters with dominant BCGs (Bautz-Morgan type I and II, Abell richness of 2 and 1), is a relatively poor cluster ($\sigma \sim 600 $km s$^{-1}$) with modest BCG (Bautz-Morgan type III).  
In this sense, these two galaxies are better local analogs of the BCGs from the lower-mass clusters of our high-$z$ sample. 
Their surface brightness profiles outward of 1 kpc from the galaxy center, as shown in Figure~\ref{fig_stack_profile_r25}, follow the high-$z$ BCG profiles and agrees particularly well with the smaller and less massive BCGs in our high-$z$ sample.
However, the best-fit \sersic parameters of M87 and M49, with half-light radius, $r_{e}=59$ kpc and 22 kpc, and \sersic index, $n=12$ and 6 \citep{Kormendy09}, have values on the large side of our BCG sample.
\citet{Kormendy09} did the \sersic fitting with their own method, rather than using GALFIT.
The systematic uncertainty caused by the different fitting methods might explain why their \sersic parameters are more similar to the high mass high-$z$ BCG while the profiles themselves resemble the lower mass high-$z$ BCGs. 
This again suggests that simple comparisons of the \sersic parameters of different studies can be dominated by systematic uncertainties in the profile-fitting procedures.

To further quantify the differences in BCG profiles shown in the above direct comparisons, we calculate the slope of the BCG profile and its central surface brightness separately and compare these quantities for BCGs of different mass and redshifts in the following sections.

\subsection{Brightness Profile Slopes in the Outer Region of BCGs}\label{s_slope}

\begin{figure*}
\epsscale{1.2}
\plotone{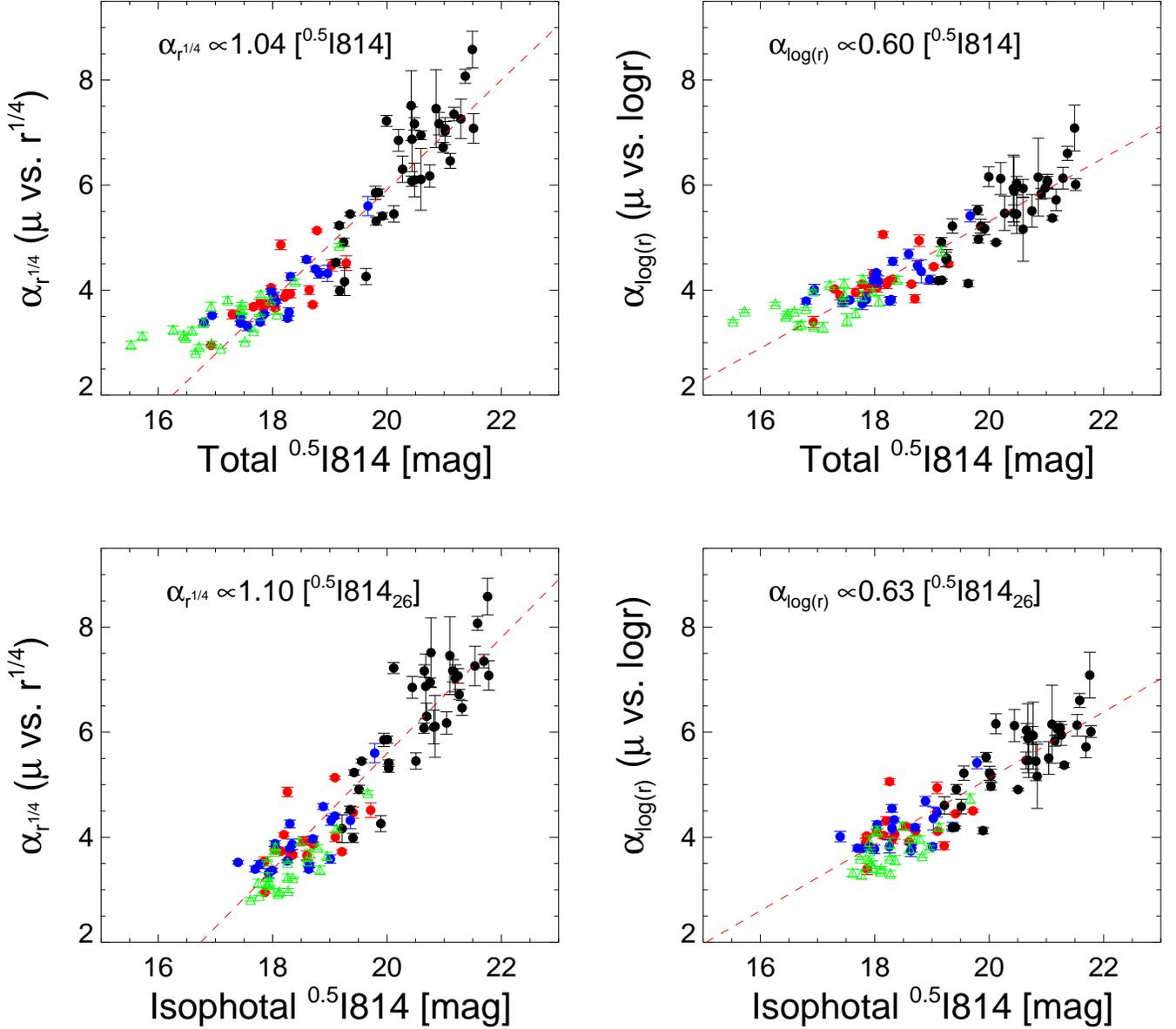}
\caption{Slopes of the brightness profiles of the BCGs and ETGs.
In the left panels, the slopes are calculated as $\mu$ vs. $r^{1/4}$, and in the right panels as $\mu$ vs. $\log r$.
In the top panels, the slopes are plotted against the total magnitude \img~integrated over the best-fit model, while in the bottom panels, the isophotal magnitudes \imgiso~ within the region that $\mu < 26$ mag arcsec$^{-2}$ are shown. 
The filled red circles are BCGs with $0.3<z<0.5$, and the blue circles are BCGs with $0.5<z<0.9$.
The black filled circles are the ETGs with $0.3<z<0.9$.
The red dashed lines are the best-fit linear correlations for the BCGs and ETGs. 
The green triangles are the local BCGs from \citet{Gonzalez05}.
}
\label{fig_slope}
\end{figure*}

To calculate the slope of the BCG profile, we perform simple linear fitting to the profile between 2 kpc from the center to the radius when $\mu$ reaches 26 mag arcsec$^{-2}$.
The inner radius cut is to ensure the slope is calculated in the region that is not affected by the PSF smoothing. 
The outer surface brightness cut, which is about 7 mag fainter than the average central brightness within 1 kpc, is to avoid the part of the profile that starts to deviate from a power-law decline, either from the sky uncertainties or the existence of fainter envelopes. 
For BCGs in our sample, their surface brightnesses reach 26 mag arcsec$^{-2}$ at around 60 kpc from the galaxy center. 
Although the actual value of the slopes does depend on the surface brightness limit adopted here, we note that the main conclusions of the paper remain the same when we vary the surface brightness limit from 25 to 27. 
For surface brightness profiles plotted as a function of $r^{1/4}$ and $\log r$, we calculate a slope for each case, $\alpha_{r^{1/4}}$ and $\alpha_{\log r}$, respectively.
We show these slopes as a function of total integrated magnitudes \img~in the top two panels of Figure~\ref{fig_slope}. 
The two types of slopes differ in value but follow a similar trend: the slope becomes shallower when the mass of the galaxy increases. 
This confirms what we see when directly comparing the profiles of galaxies.
The dependence of the $r^{1/4}$ slope on the total magnitude is slightly steeper than the logarithm slope.
The dependence of the total magnitude on the slope is tighter for $\alpha_{r^{1/4}}$ than for $\alpha_{\log r}$ (a scatter of 0.5 mag versus a scatter of 0.6 mag).
This is because the surface brightness profile plotted in $\log r$ deviates more from a linear decline compared to the profile plotted in $r^{1/4}$, and therefore $\alpha_{\log r}$ is a poorer decline rate indicator than $\alpha_{r^{1/4}}$. 

For both slopes, the dependence also appears to be steeper for the less massive ETGs and becomes flatter for the more massive BCGs.
The more massive BCGs have more extended outer envelopes ($\sim 100$ kpc) that add a lot of light to the galaxy without changing the inner profile too much. 
This can produce a large spread in total magnitudes while keeping inner profiles the same. 
Because the systematics in the sky measurement can greatly affect the detection of the outer envelope, it could affect the relation between the profile slope and the measured total magnitude.
To minimize the dependence on the sky measurements, instead of using total integrated magnitudes, we measure the isophotal magnitudes \imgiso, which is the total flux enclosed in the region of galaxy that is brighter than 26 mag arcsec$^{-2}$.
This isophotal magnitude is more consistent with the slope we measure because they are both measured within the same isophotal radius.
In the bottom two panels of Figure~\ref{fig_slope}, we plot the profile slopes versus the isophotal magnitudes.
In both plots, the profile slopes show slightly steeper relations with the isophotal magnitudes than with the total magnitudes, but not by much. 

In general, the profile slopes increase with the magnitudes as $\alpha_{r^{1/4}}\propto$\img~ and $\alpha_{\log r}\propto0.6$\img, for both total and isophotal magnitudes.
If we convert the magnitudes to stellar mass, the correlations translate into $\alpha_{r^{1/4}}\propto M_{*}^{-2.5}$ and $\alpha_{\log r}\propto M_{*}^{-1.5}$.

\subsection{Central Brightness of BCGs}\label{sec_cb}
\begin{figure}
\epsscale{1.2}
\plotone{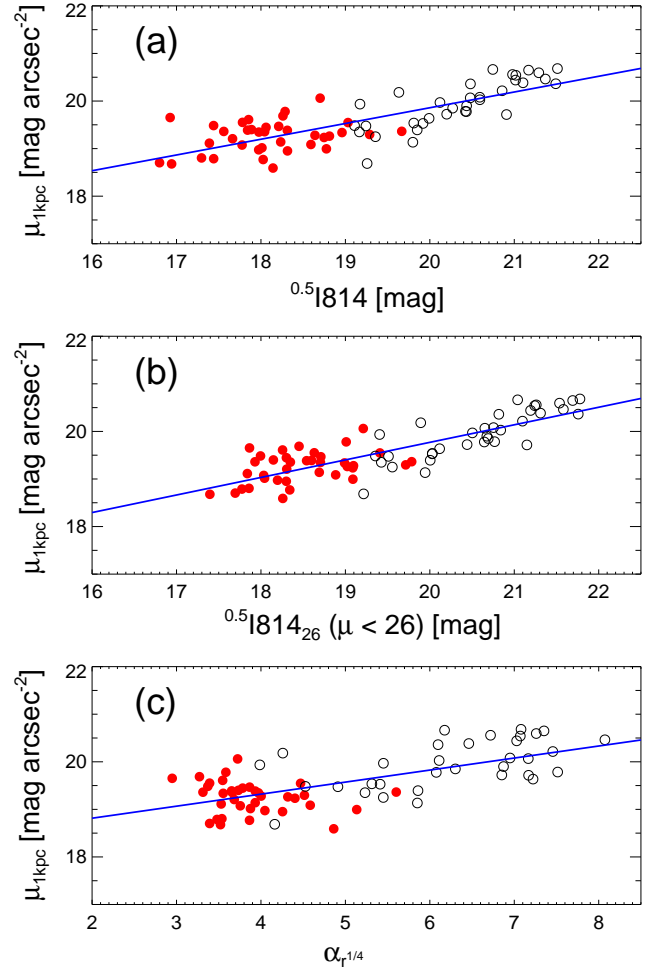}
\caption{Surface brightness at 1 kpc radius of the BCGs and ETGs as a function of (a) total integrated magnitude of the best-fit \sersic model \img; (b) isophotal magnitudes down to $\mu < 26$ mag arcsec$^{-2}$; (c) the slope $\alpha_{r^{1/4}}$ of the brightness profiles in the outer regions.
The red filled circles are the BCGs, and the open black circles are the ETGs.
The blue line in each panel is the best-fit linear correlation for both BCGs and ETGs.
}
\label{fig_mu0}
\end{figure}

Direct comparison of the BCG surface brightness profiles shows the central brightness of the BCGs spans a small range. 
The surface brightnesses at 1 kpc from the galaxy centers, $\mu_{\rm 1kpc}$, have an average value of 19.25 mag arcsec$^{-2}$ with a scatter of 0.33, much smaller than the scatter of $0.67$ in their total magnitudes. 
Because the FWHM of the PSF of our data has a typical physical size of $\sim$0.9 kpc, $\mu_{\rm 1kpc}$ correlates very well with the average surface brightness within 1 kpc, only fainter by 0.07.
In panel (a) of Figure~\ref{fig_mu0}, we plot $\mu_{\rm 1kpc}$ of the BCGs and the ETGs as a function of their total magnitudes integrated over the best-fit \sersic model. 
If we look at the BCGs alone, $\mu_{\rm 1kpc}$ is a pretty flat function of the total magnitude.
A Spearman's test does not reject the null hypothesis of no correlation with with a $p$-value of 0.22. 
But if we combine BCGs and ETGs together and probe a larger mass range, $\mu_{\rm 1kpc}$ does show as a slow function of the total magnitude that the brighter and bigger BCGs/ETGs have slightly brighter central surface brightness, with $\mu_{\rm 1kpc} \propto 0.35~$\img. 
The Spearman's coefficient is 0.64, and the null hypothesis of no correlation is rejected with a $p$-value of $2\times10^{-7}$.

As discussed before, the total magnitude of the best-fit \sersic model is sensitive to the uncertainties in the sky background measurement and the contamination from ICL.
The large spread in the total magnitudes could therefore partly come from these factors that are not directly related to the central brightness of the BCGs and weaken any possible correlations between them. 
The isophotal magnitude, \imgiso, is more robust against these complications and should correlate better with the slope that we measure within the same isophotal radius.
As shown in panel (b) of Figure~\ref{fig_mu0}, this does indeed strengthen the correlation.
For the BCGs, the Spearman's coefficient is 0.38, and the null hypothesis of no correlation is rejected with a $p$-value of 0.02.
Their central brightness increases as $\mu_{\rm 1kpc} \propto 0.25\mbox{\imgiso}$. 
For the BCGs and ETGs together, the correlation coefficient is 0.7 with a P-value of $2\times10^{-9}$ to reject the null hypothesis of no correlation.
In this case, the growth is also faster, with $\mu_{\rm 1kpc} \propto 0.39$\imgiso.
If the mass-to-light ratio of a BCG/ETG is the same throughout the whole galaxy, these relations imply the central mass density within 1 kpc of the galaxy increases with the mass of the galaxy as $\rho_{\rm 1kpc}\sim M_{*}^{0.25}$ for the BCGs alone and as $\rho_{\rm 1kpc}\sim M_{*}^{0.39}$ for the BCGs and ETGs together.  
These relations are shallower than what \citet{Saracco12} found for ETGs, $\rho_{\rm 1kpc}\sim M_{*}^{0.6}$, but ETGs in their sample are generally less massive ($M_{*}<3\times10^{11} M_{\sun}$) than the BCG/ETG sample we have here. 
The supermassive ETGs studied by \citet{Tiret11} have masses more comparable to our sample, $M_{*}\sim10^{11}-5\times10^{12} M_{\sun}$, but the relation they found, $\rho_{\rm 1kpc}\sim M_{*}^{-0.25}$, has opposite trend compared to other studies.
\citet{Saracco12} argue the difference is due to how the mass is estimated in different studies.
\citet{Tiret11} estimate the dynamical mass by solving the Jeans equation utilizing surface brightness profile and central velocity dispersion, while \citet{Saracco12} use the light profile as the tracer of the mass assuming a fixed mass-to-light ratio, which is the method used in this paper.  
In fact, \citet{Saracco12} have shown that if they apply the same mass estimate to the supermassive ETGs from \citet{Tiret11}, they recover a similar positive trend between $\rho_{\rm 1kpc}$ and $M_{*}$ and the trend appears to be shallower than that of the less massive ETGs (see left panel of Figure~$7$ in their paper), which would be more consistent with the trend we find here.

We also look into the dependence of the central brightness on the outer profile slopes, as shown in panel (c) of Figure~\ref{fig_mu0}.
For BCGs, the central brightness $\mu_{\rm 1kpc}$ does not show obvious dependence on the outer profile slope $\alpha_{r^{1/4}}$. 
The Spearman's coefficient is $-0.27$ with a $p$-value of 0.11 for no correlation.
If we consider the BCGs and ETGs together, we have a correlation $\mu_{\rm 1kpc} \propto 0.28\alpha_{r^{1/4}}$ with a Spearman's coefficient of 0.41 and a $p$-value of 0.002.
The lack of correlation between $\mu_{\rm 1kpc}$ and $\alpha_{r^{1/4}}$ of BCGs suggests the growth of the inner core of BCGs is largely decoupled from the growth of the outer region.

\subsection{BCG Structural Parameters and Host Clusters}
The properties of BCGs have been shown to correlate positively with their host clusters in the sense that the more massive clusters tend to host more massive BCGs \citep[e.g.,][]{Lin04,Whiley08,Stott10,Lidman12,Stott12}. 
For the BCGs from the RCS sample, the cluster-galaxy correlation amplitude, $B_{\rm gc}$, is a robust indicator of the richness of the cluster \citep{Yee99}.
It has been shown to have a strong correlation with the velocity dispersion of clusters and can be used as a cluster mass estimator \citep{Yee03, Muzzin07}.

\begin{figure*}
\epsscale{1.2}
\plotone{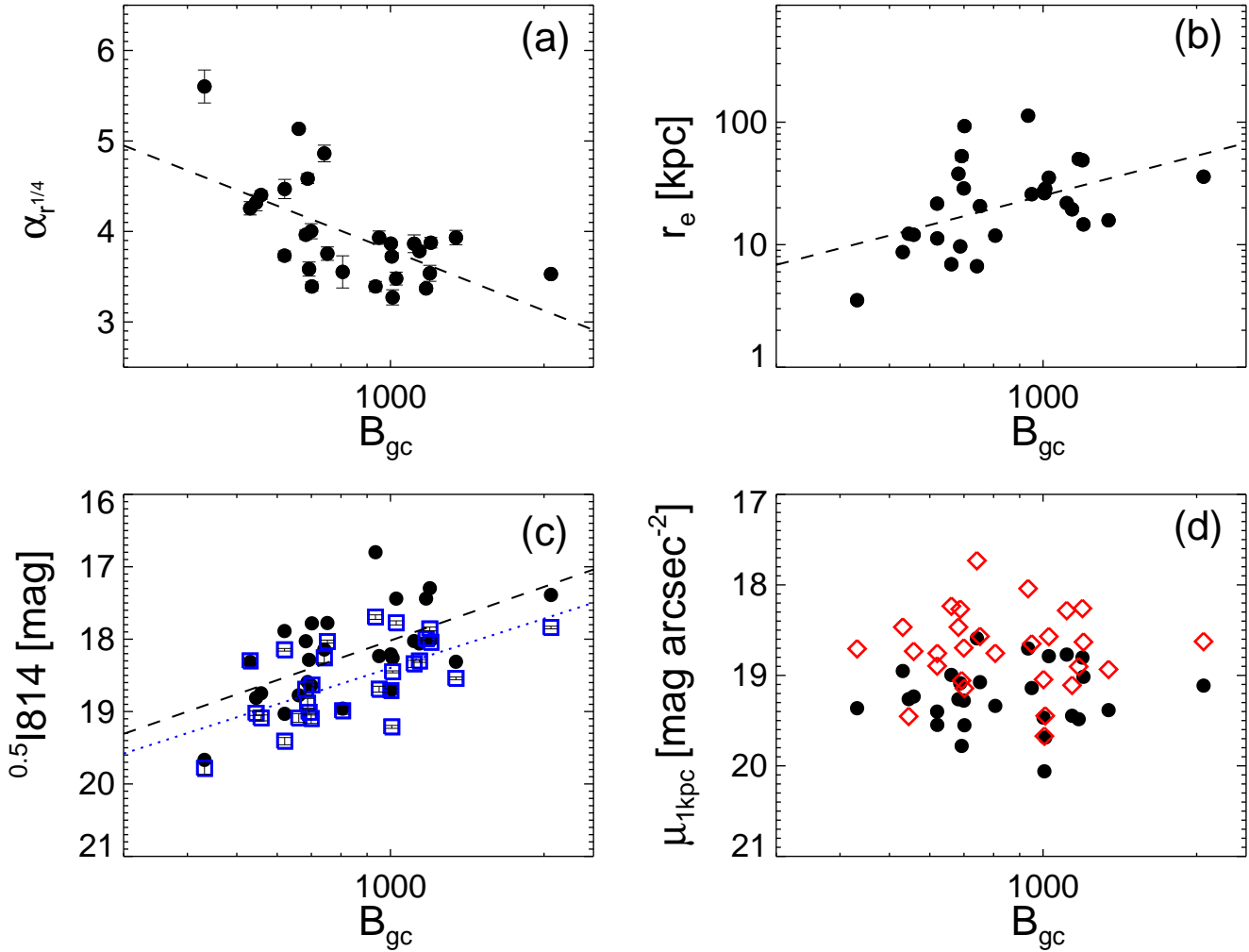}
\caption{BCG properties as a function of the host cluster richness parameter $B_{\rm gc}$.
(a) Slope of the outer brightness profile; (b) effective radius of the best-fit \sersic model; (c) total magnitude integrated over the best-fit \sersic model (black filled circles) and the isophotal magnitude down to $\mu<26$ mag arcsec$^{-2}$ (blue open squares); and (d) measured surface brightness at 1 kpc (the filled black circles) and the intrinsic average surface brightness within 1 kpc (red open diamonds).
In panels (a)-(c), the dashed lines are the linear fits of the black data points.
The blue dotted line in panel (c) is the linear fit for the isophotal magnitude (blue open squares) as a function of $B_{\rm gc}$.
}

\label{fig_bgc}
\end{figure*}

In Figure~\ref{fig_bgc}, we show the properties of the BCGs as a function of the richness parameter $B_{\rm gc}$ of their host clusters.
In these plots, we include all RCS BCGs except RCS$1319-02$ whose $B_{\rm gc}$ value is very small and has a large uncertainty.
Among the different properties, both the outer profile slopes, $\alpha_{r^{1/4}}$, and the total integrated magnitudes correlate strongly with the cluster richness. 
The Spearman test rejects the null hypothesis of no correlation with a $p$-value $<0.001$.
The correlation between $r_{e}$ and $B_{\rm gc}$, on the other hand, is weaker, with Spearman's coefficient of 0.4 and $p$-value$=0.016$.
More specifically, the slopes of the outer profile of the BCGs become shallower with the increase of cluster richness, $\alpha_{r^{1/4}}\propto (-2.2\pm0.6)\log B_{\rm gc}$ and $r_{e}$ becomes larger, $\log(r_{e})\propto (1.1\pm0.4)\log B_{\rm gc}$.
The total magnitudes of the BCGs change as \img$\propto (-2.5\pm0.6)\log B_{\rm gc}$.
As we discussed earlier, the total integrated magnitude \img~is sensitive to the systematic uncertainties and the ICL contamination. 
Therefore we also plot the isophotal magnitude \imgiso~as a function of $B_{\rm gc}$ in panel (c) of Figure~\ref{fig_bgc}.
It gives a similar correlation \imgiso$\propto (-2.3\pm0.6)\log B_{\rm gc}$ with a slightly smaller scatter ($\sigma = 0.45$ versus $\sigma = 0.50$).

Because the cluster richness parameter $B_{\rm gc}$ scales with the cluster mass as $\propto M^{0.6}_{\rm cluster}$ \citep{Yee03}, the relations we find between the magnitudes of BCGs and $B_{\rm gc}$ would imply that the stellar mass of BCGs, both the total mass and the isophotal mass, scale with the cluster mass as 
\begin{equation}
M^{*}_{\rm BCG}\propto M^{0.6\pm0.1}_{\rm cluster}
\end{equation}
This relation is steeper than the index of $\sim0.26\pm0.04$ found by \citet{Lin04} but agrees very well with the index $\sim0.6\pm0.1$ found by \citet{Lidman12}.

The central surface brightnesses of the BCGs, neither the apparent measured value nor the intrinsic value deduced from the \sersic fitting, show any correlation with the richness of the clusters (see panel (d) of Figure~\ref{fig_bgc}).
Spearman's rank tests suggest that both relations have more than $75\%$ probability of being random correlation. 

The lack of correlation between the central surface brightness of BCGs and the cluster mass suggests the correlation between the cluster mass and BCG mass is driven by growth in the outer part of the BCGs.

\section{DISCUSSION}

\subsection{Kormendy Relation}
\begin{figure}
\epsscale{1.2}
\plotone{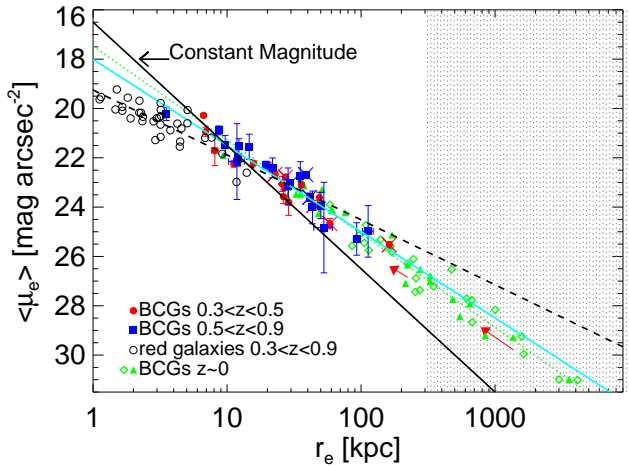}
\caption{Kormendy relation for the BCGs with single \sersic profile fits. The red filled circles are the BCGs with $0.3<z<0.5$, and the blue squares are the BCGs with $0.5<z<0.9$. The ones with crosses are the BCGs from the X-ray cluster sample. The green filled triangles are the local BCGs from \citet{Gonzalez05} ($z\sim0.1$), and the green open diamonds are from \citet{Brown97} ($z<0.1$). 
The ETGs with spectroscopic redshifts $0.3<z<0.9$ are shown as black open circles.
The cyan line is the best fit to our BCG sample, and the green dotted line is for local BCGs. The black dashed line is the best fit for the lower-mass ETGs.
Two red arrows show how the systematics in the profile measurement move two local BCGs to smaller sizes. The arrows have been offset downward by 0.5 on the $y$ axis for clarity.
The shaded region indicates a size that is larger than the BCG images used in this study. 
The black solid line shows the slope of the constant magnitude relation as a reference.
}
\label{fig_kmd_sers}
\end{figure}

The Kormendy relation is the relation between the size and the average surface brightness of ETGs.
It has been regarded as one of the fundamental scaling relations for these galaxies and is used widely as a tool to analyze surface brightness profile evolution. 
With the best-fit \sersic model, we can construct the Kormendy relation for our BCG sample using the half-light radius, $r_{e}$, along the major axis, a measure of BCG size, and $\langle\mu_{e}\rangle$, the average surface brightness within $r_{e}$, as the BCG's average surface brightness.
We show this relation for our BCG sample in Figure~\ref{fig_kmd_sers}.
Again, all the quantities shown in the figure have been shifted to their corresponding values at redshift $0.5$. 
 
In general, the BCGs form a fairly tight Kormendy relation, which can be described by
\begin{equation}
\langle \mu_{e} \rangle = (18.01\pm0.23) + (3.50\pm0.18) \log(r_{e})  
\end{equation}
The scatter in the relation is 0.41 dex, and it is steeper than the relation formed by less massive ETGs.
The Kormendy relation of the ETGs has a slope of $2.63\pm0.28$, which is quantitatively consistent with the previous findings \citep{Bildfell08}.

We note that both of these relations are shallower than a relation between the mean surface brightness and the half-light radius for any profiles with a constant magnitude. 
By the simple definition of the mean surface brightness and the half-light radius, such a relation will always have a slope of 5 (shown as the black line in Figure \ref{fig_kmd_sers}).
In other words, a population of galaxies with constant magnitudes will always form a linear correlation on the $\langle \mu_{e} \rangle$ versus $\log r_{e}$ plane with a slope of 5 regardless of the actual profiles the individual galaxies follow, which may not even be anything like a \sersic profile. 
The fact that our BCG sample spreads across two magnitudes and formed a tight Kormendy relation with a shallower slope clearly shows that their Kormendy relation is not simply a convergence onto a constant magnitude relation.

If we split our BCG sample into two redshift bins, $0.3<z<0.5$ and $0.5<z<0.9$, we find no significant difference in their Kormendy relations: neither in the slope, the normalization, nor the range of sizes they span.
This seems to agree with the minimal evolution in the Kormendy relation found by \citet{Stott11}.

Using the single \sersic measurements of the local BCGs from \citet{Gonzalez05} ($z\sim0.1$) and \citet{Brown97} ($z<0.1$), we can also plot their Kormendy relation in the same figure.
Despite the very different size range of local BCGs and our BCG samples, the local BCG Kormendy relation has a very similar slope and normalization as the relation of our BCG sample. 
In the size range of 20--100 kpc, the two relations overlap with each other.  
As discussed in \S \ref{sec_issue}, the systematic errors in the sky background measurement can affect the actual value of the \sersic parameters, and this may affect the comparison of the Kormendy relations. 
Using the simulations done in \S \ref{sec_issue}, we can show (red arrows in Figure \ref{fig_kmd_sers}) how the estimates of $r_{e}$ and $\langle\mu_{e}\rangle$ of two local BCGs, A2969 and A2730 from \citet{Gonzalez05}, would change if we observe them at redshift $0.5$ and deduce their profile parameters with the same type of data as our BCG sample.
It is clear that the systematics in the parameter measurement due to the sky background measurement conspire to move the BCGs along their Kormendy relation: although the values of the individual parameters do change a lot, the changes are somehow coupled to roughly conserve the slope of the Kormendy relation.      
This result has been reported before \citep[e.g.,][]{Gonzalez05}, and its direct implication is that the slope of the Kormendy relation is insensitive to the systematics in the \sersic profile fitting, and the similarity in the slopes of the Kormendy relation between our BCG and local BCG sample is a robust result.

Not only is the Kormendy relation insensitive to the systematic error in the sky background measurement, it is also insensitive to the different \sersic indices used in the profile fitting.
Although a single \dev model tends to fall short of reproducing the observed profile at the large radii and hence yields a smaller $r_{e}$, the Kormendy relation generated from the best-fit parameters of a \dev model has almost identical normalization and slope as that given by the \sersic model, except that all the data points now move to smaller radii.

\begin{figure}
\epsscale{1.2}
\plotone{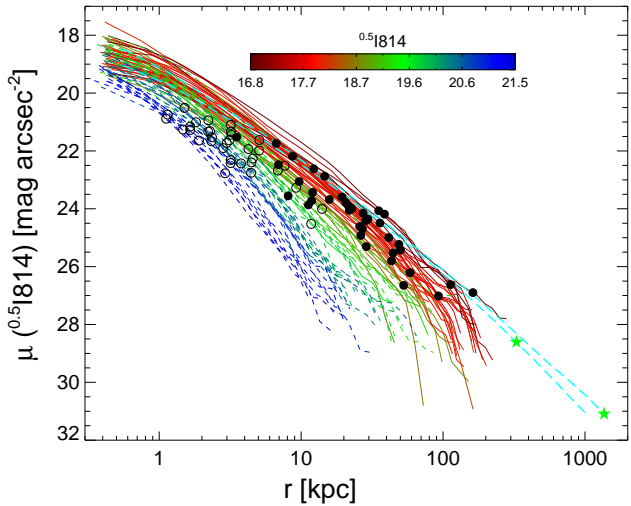}
\caption{Kormendy relation for BCGs shown as $\mu_{e}$, the surface brightness at $r_{e}$, as a function of $r_{e}$.
The filled black circles are for BCGs, and the open ones are for ETGs.
Along with the Kormendy relation, the surface brightness profile of each galaxy is also shown as a colored curve.
The solid profiles are BCGs, and the dashed ones are ETGs.
The profiles are color-coded by the integrated \img~of the best-fit \sersic model, as shown by the color bar.
The two cyan curves are the best-fit models for two local BCGs, A2629 and A2730, from \citet{Gonzalez05}. 
The green stars indicate the location of $r_{e}$ for these two galaxies.
}
\label{fig_stack_profile_logr}
\end{figure}

Why does the Kormendy relation of BCGs barely change, even though the individual model parameters change substantially due to the systematics?
Why does a \dev model, which is inadequate to describe the real profile, still give parameters that fall on the same Kormendy relation?
To understand this, we can show the Kormendy relation in a slightly different form.
Instead of $\langle\mu_{e}\rangle$, the mean surface brightness within $r_{e}$, we can plot $\mu_{e}$, the surface brightness at $r_{e}$, as a function of $r_{e}$ (Figure. \ref{fig_stack_profile_logr}).
Both forms of the Kormendy relation have been widely used in the literature, and the two relationships only differ by a constant in normalization if galaxies have \sersic profiles with the same \sersic index \citep{Graham97}.
For \sersic profiles with different indices, the difference between $\langle \mu_{e} \rangle$ and $\mu_{e}$ depends only weakly on $n$, $\mu_{e} - \langle \mu_{e} \rangle \propto 1.2\log n$. 
Although the actual galaxy profiles deviate from perfect \sersic profiles, $\langle \mu_{e} \rangle$ and $\mu_{e}$ calculated directly from the observed profiles of our galaxies are still well correlated with a constant difference of 1.4 mag arcsec$^{-2}$ with a small scatter of 0.18.

With $\mu_{e}$ instead of $\langle \mu_{e} \rangle$, we show that the Kormendy relation overlaid directly on the surface brightness profiles of BCGs, as in Figure~\ref{fig_stack_profile_logr}. 
With this, it becomes clear that the Kormendy relation of the BCGs traces the average slope of the BCG profiles because these profiles form a close ``bundle". 
As discussed before, the systematics in the sky background measurement can strongly affect the value of $r_{e}$.
However, as long as it does not change the slope of the surface brightness profiles too much, the slope of the Kormendy relation will not change much neither.
Instead, the parameters of the galaxies will only move along the Kormendy relation.
This explains why different studies of BCG profiles give controversial results on the size evolution of the BCGs but generally agree on the nonevolution of the slope of the Kormendy relation. 
This also explains why the single \dev profile fit of our BCG sample, though generally inferior to the \sersic profile fit and producing smaller $r_{e}$, only shifts galaxy along the same Kormendy relation.

\subsection{Evolution of BCG Structure Parameters}
\subsubsection{Evolution of the Central Surface Brightness}
\begin{figure}
\epsscale{1.2}
\plotone{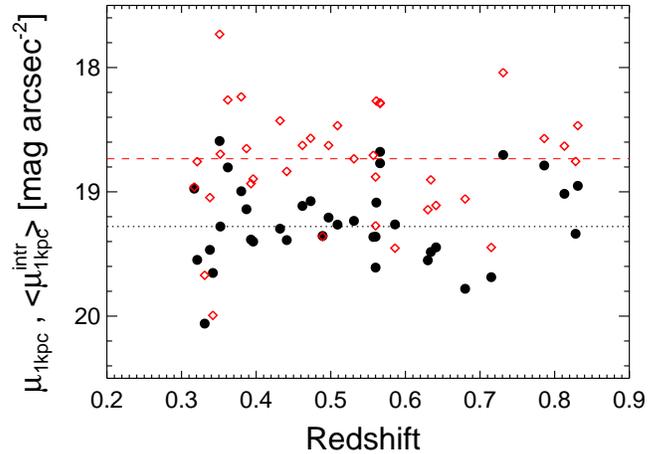}
\caption{Central surface brightness within 1 kpc of a BCG's center as a function of redshift.
The red open stars are the intrinsic average brightness within 1 kpc deduced from the best-fit \sersic profiles, and the black filled circles are the direct measurement from the profiles at 1 kpc.
The red dashed line and the black dotted line are the median values of each measurement.
}
\label{fig_mu0_z}
\end{figure}

An almost unchanged central brightness of the ETGs across large ranges of total luminosity and redshift has already been reported by many studies \citep[e.g.,][]{Bezanson09,vanDokkum10,Tiret11,Patel13,Saracco12}.
In particular, \citet{Tiret11} has reported the central stellar mass density of a sample of $z<0.3$ supermassive ETGs remains almost constant $\sim2\times10^{10} M_{\sun}~ \mbox{kpc}^{-3}$ and that it is not very different from the ETGs at $z\sim2$.
As shown in \S \ref{sec_cb}, the central surface brightness at 1 kpc ($\mu_{\rm 1kpc}$) of our measured brightness profile is almost identical to the average surface brightness within 1 kpc. 
However, because of the PSF smoothing of the profile, $\mu_{\rm 1kpc}$ is different from the intrinsic, unsmoothed average surface brightness within 1 kpc, $\langle \mu_{\rm 1kpc}^{intr} \rangle$. 
The latter can be estimated from the best-fit \sersic models.
In Figure~\ref{fig_mu0_z}, we show both $\mu_{\rm 1kpc}$ and $\langle \mu_{\rm 1kpc}^{\rm intr} \rangle$ as a function of redshifts. 
The value of $\langle \mu_{\rm 1kpc}^{\rm intr} \rangle$ is on average $\sim0.5$ mag brighter than $\mu_{\rm 1kpc}$.
Neither of the two central surface brightnesses of the BCGs show a dependence on redshifts.

From the intrinsic central surface brightness, we can estimate the total luminosity within 1 kpc radius, $L_{<\rm 1kpc}$, and relate it to the stellar mass, $M_{*<\rm 1kpc}$, using the mass-to-light ratio given by the SED model of \citet{Maraston09}.
With a sample of ETGs at $0.9<z<2$, \citet{Saracco12} estimated the central stellar mass density by assuming $\langle \rho_{\rm 1kpc} \rangle=4/3\pi M_{*<\rm 1kpc}$ and found a correlation between $\langle \rho_{\rm 1kpc} \rangle$ and the total stellar mass of ETGs, independent of redshift.
Following their definition, we find the BCGs in our sample have $\log \langle \rho_{\rm 1kpc} \rangle = 10.20\pm0.18 M_{\sun}~ \mbox{kpc}^{-3}$. 
This value agrees very well with the central stellar density at total stellar mass, $M_{*}\sim10^{12} M_{\sun}$, extrapolated from the relation given by \citet{Saracco12}.
It also agrees well with the central stellar mass density of the supermassive ETGs at $z<0.3$ from \citet{Tiret11}.
This further reinforces the conclusion that the central mass densities of the massive ETGs show little evolution across a large redshift range.

We note this central mass density estimate does not take into account the projection effect because it assumes all the mass within the projected 1 kpc radius is enclosed in a 1 kpc radius sphere. 
We can estimate the difference between $\langle \rho_{\rm 1kpc} \rangle$ and the true central mass density using the Prugniel--Simien density model \citep{Prugniel97}, which is designed to match the deprojected form of the \sersic model \citep{Graham06}.
The true, deprojected central mass density is about 0.2 dex smaller than the projected one. 
This offset is almost the same for all the ETGs studied here, and it does not affect the results of the comparisons.

\subsubsection{Evolution of the Slope}
The two subsamples of our high-$z$ BCGs, with $0.3<z<0.5$ and $0.5<z<0.9$, show no distinguishable difference in their outer profile slopes; see Figure~\ref{fig_slope}.
Not only do the slopes of the two subsamples follow the same correlation with magnitudes, but they also span the similar ranges in the two parameter spaces.

To compare the slopes of the profiles of high-$z$ BCGs with local BCGs, we adopt the best-fit \sersic profiles for the BCGs from \citet{Gonzalez05} ($z\sim0.1$) and calculate the slopes in the same fashion as for our high-$z$ BCGs.
We note that although we do not derive the slopes directly from the surface brightness profiles as we do for the high-$z$ sample, the best-fit \sersic profiles of the local BCGs match the real profiles very well within the region in which we calculate the slopes ($r>2$ kpc and $\mu_{^{0.5}I814} < 26$ mag arcsec$^{-2}$).  
In the top panels of Figure~\ref{fig_slope}, the profile slopes of the local BCGs continue the same trend between the slope and the total magnitude of high-$z$ BCGs but with brighter luminosities.
The flattening of the slope at the bright end becomes even more pronounced.
This reinforces the argument that the increase in the total magnitudes of these BCGs is driven by the faint envelopes (or induced by systematics) beyond the isophotal radius $r_{\mu<26}$. 

If we plot the slopes of the local BCGs as a function of the isophotal magnitudes, as shown in the two bottom panels of Figure~\ref{fig_slope}, then they follow the same relation defined by the high-$z$ BCGs and the difference in the magnitudes between local BCGs and high-$z$ BCGs becomes less significant.
However, the local BCGs do appear to have slightly shallower slopes than high-$z$ BCGs.
The robust mean of $\alpha_{r^{1/4}}$ for the local BCG is $3.37\pm0.08$, compared to $3.81\pm0.07$ for the high-$z$ BCGs.
The difference is partly due to the fact that the BCGs in the local sample are generally more massive than the ones in the high-$z$ sample.
If we limit the comparison to the BCGs with $18<$\imgiso$<19$, the difference becomes smaller, $3.59\pm0.20$ versus $3.76\pm0.05$.

As we discussed in the last section, BCG profiles also correlate with the masses of host clusters: the more massive the cluster is, the more massive the BCG is, and the shallower the slope.
The bias in the cluster sample selection can therefore affect the direct comparison of the slopes.
To track the evolution of BCGs properly, we need to take into account the difference in the cluster mass of different samples.
For this purpose, we construct a ``refined" high-$z$ BCGs sample by only selecting BCGs hosted by clusters with $B_{\rm gc}>800$. 
This includes 14 BCGs with their host clusters with $B_{\rm gc}$ ranging from 800 to $\sim2000$.
\citet{Yee03} have shown that $B_{\rm gc}$ scales well with cluster mass as $B_{\rm gc}\propto M^{0.6}$, and it can be used to estimate the cluster mass reliably. 
Using this scaling relation, the refined high-$z$ BCG sample has a mass range of $5$-$20\times10^{14} M_{\sun}$ and an average mass of $7\times10^{14} M_{\sun}$. 
For the local BCGs, we do not have the $B_{\rm gc}$ measurements for their host clusters, but \citet{Zaritsky06} have measured the velocity dispersions of these clusters, and we can use them to estimate the cluster mass.
The velocity dispersions of the local cluster sample have a large spread and there are eight of them with velocity dispersion greater than $800$ km s$^{-1}$.
Among these eight clusters, two, A3693 and A3705, have been shown to have significant subclumps which might result in an overestimated velocity dispersion \citep{Sivanandam09}.  
We therefore exclude these two clusters and construct the refined local sample with six clusters. 
These six clusters have velocity dispersion ranging between $840$-$1050$ km s$^{-1}$, which corresponds to a cluster mass range of $10$-$19\times10^{14} M_{\sun}$.
The average cluster mass of the sample is $14\times10^{14} M_{\sun}$. 
From the refined $z=0.5$ BCG sample to the refined local BCG sample, the cluster masses have grown by a factor of two. 
Because the mass growth rate of clusters from $z=0.5$ to $z=0.1$ is also expected to be about a factor of two \citep[e.g.,][]{Fakhouri10},  it means the clusters in our refined high-$z$ sample will eventually grow into clusters in the refined local sample.  

The robust mean of the slope for the refined local sample is $3.01\pm0.07$, which is much shallower than the mean of the high-$z$ refined sample, $3.62\pm0.07$.
Therefore, for the clusters that have grown their mass by a factor of two from $z=0.5$ to $z=0.1$, the growth rate expected by the $\Lambda$CDM model, their BCG profile slopes have decreased by $\sim0.6$.

\citet{Stott11} compared five high-$z$ BCGs (z$\sim$1) to 19 BCGs at $z\sim0.2$ and found no significant differences in their Kormendy relations and the range of $r_{e}$ that two samples have. 
Prompted by these findings, they concluded that there is no evolution in the BCG profiles from $z\sim1$ to $z=0.2$. 
Using the stacked profiles of the high-$z$ and low-$z$ BCG samples provided in the same paper, we can calculate the average $\alpha_{r^{1/4}}$ slope for these two samples.  
Their low-z BCG sample has redshift only slightly higher than the local BCGs from \citet{Gonzalez05} (0.2 versus 0.1), and their average slope, $\alpha_{r^{1/4}}$=3.2, is similar to the average slope (3.37) of the local BCG sample from \citet{Gonzalez05}.
Their high-$z$ BCGs, which are at higher redshift than our BCGs sample (1 versus 0.5), have an average slope, $\alpha_{r^{1/4}}$=3.5, more similar to the mean of our refined high-$z$ sample than the mean of the whole high-$z$ sample (3.6 and 3.8).
The five high-$z$ BCGs in \citet{Stott11} are all from X-ray luminous clusters, with three of them having $L_{x}>10^{45} \mbox{erg~s}^{-1}$ \citep{Cavagnolo08}, while the RCS clusters typically have $L_{x}<10^{45} \mbox{erg~s}^{-1}$ \citep{Hicks08}.
This suggests that the high-$z$ clusters in \citet{Stott11} are likely more massive on average than the typical RCS clusters, and this might explain why their slope matches our refined high-$z$ sample better.
Despite all the subtle differences, the profiles from \citet{Stott11} generally confirm the evolution of the BCG profiles found between our high-$z$ BCG sample and the local sample from \citet{Gonzalez05}.  

\subsubsection{Evolution of the Stellar Mass}
In addition to the profile slope, we can also compare the stellar mass of the high-$z$ BCGs to the local BCGs. 
For the following comparisons, we only consider the refined samples that have taken into account the cluster evolution.
The average total magnitudes \img~ of the BCGs in the local refined sample and the high-$z$ refined sample are $16.70\pm0.26$ and $17.94\pm0.16$ mag, respectively.
This corresponds to a factor of three in the mass growth rate of BCGs. 
However, as we discussed earlier, the total integrated magnitudes derived from the best-fit \sersic profile depend sensitively on the outer faint envelope of the galaxy and therefore suffer from systematics in the sky background measurements. 
This should be less of a problem for comparison within the same data set but could be a serious issue if comparing results from different sets of data.
This makes it hard to accurately estimate the evolution of the BCG stellar mass.

To minimize the systematics between high-$z$ and local BCG samples, we limit the comparison to isophotal magnitudes instead of total magnitudes. 
The local refined sample and high-$z$ refined sample have average isophotal magnitudes, \imgiso~ of $17.9\pm0.1$ and $18.3\pm0.1$, respectively. 
This would imply a mass growth rate of $1.5$ in the BCG isophotal stellar mass from $z=0.5$ to $z=0.1$, while their host cluster masses grow by a factor of two.
Because the isophotal magnitude/stellar mass is limited within the isophotal radius of the BCGs, it does not represent the complete picture of the mass growth of the BCGs.
The isophotal mass growth we report here should be taken as the lower limit of the total mass growth of the BCGs. 

A lower limit of $1.5$ in the mass growth rate of BCGs agrees reasonably well with the factor of two predicted from simulations.
Using NIR photometry, \citet{Lidman12} reported a factor of 1.8 mass growth for BCGs from $z=0.9$ to $z=0.2$ after matching cluster masses at different redshifts to take into account the cluster mass growth.
In general, their result agrees quite well with what we find here, but with a slightly larger difference from the model predictions.

\subsection{Inside-Out Growth of BCGs}
The central surface brightnesses within 1 kpc of the BCGs in our sample span a small range, with a scatter of $\sim0.33$ mag, only about half of the scatter in their total magnitudes. 
The corresponding central stellar mass density of the BCGs shows a weak dependence on the mass of galaxy, and it shows no strong evolution in the redshift range $0.3<z<0.9$.
Furthermore, it shows no dependence on the mass of the host cluster. 
The surface brightness profile slope at $r>1$ kpc, on the other hand, correlates strongly with the total mass of the BCGs and the mass of the host cluster.
Such characteristics of the BCG profile strongly support inside-out mass growth, in which the central part of the galaxy forms first, and the following growth mainly happens in the outer regions.

This inside-out pattern of growth is consistent with the results from studies of massive ETGs through a large redshift range 
\citep{Hopkins09,Bezanson09,vanDokkum10,Patel13}. 
In particular, \citet{vanDokkum10} tracked a population of massive galaxies with constant 
number density from $z=2$ to $z=0.6$ and studied the evolution in the average surface brightness profiles.
At $z<1$, the population of massive galaxies in their study are predominantly red passive galaxies with average mass of $\log M_{*}/M_{\sun}=11.35$ at $z=0.6$.
This is about the stellar mass range of the ETGs in our sample.
This agrees very well with our finding here, but we note that our result is based on comparing massive galaxies in the same epoch with different masses, while their result is based on comparing the same population of galaxies at different redshifts.

Because the stellar population of BCGs is predominantly an old population formed at $z>2.5$ \citep[e.g.,][]{Thomas05,Jimenez07}, and they are expected to be assembled hierarchically through mergers of smaller galaxies at a late time in the $\Lambda$CDM model, the mass growth of the BCGs at $z<2$ is expected to be dominated by dry mergers \citep{deLucia07}.
Theoretically, late-time dry mergers of the massive ETGs with less massive galaxies are expected to grow the outer regions of the galaxies while leaving the central dense region unchanged \citep[e.g.,][]{Hopkins09,Naab09,Taranu13}. 
This expectation agrees with our findings very well and supports a connection between the inside-out growth and dry mergers.
However, from our data alone, it is not clear if this growth pattern is truly unique to dry mergers.
Because our galaxy sample does not include galaxies that might grow mostly through gas-rich mergers, we cannot exclude the possibility that gas-rich mergers can also produce profiles that follow such a pattern of growth.

Interestingly, \citet{Saracco12} have also reported on a small scatter in the central surface brightness of a sample of high-$z$ ETGs ($0.9<z<2$) relative to the large span in their effective mass densities. 
However, they argued that any light profile that follows a \sersic profile with loosely constrained parameters, i.e., $n\sim4$, $0.5<r_{e}<10$ kpc and a total stellar mass $10^{10}$ to $4\times10^{11} M_{\sun}$, will span a small range in the central surface brightness and a much large range in the effective mass density. 
Hence, they concluded that the observed small scatter in the central surface brightness of ETGs is simply a peculiar feature of the \sersic profile and has no connection with inside-out growth. 
Although the BCG profiles in this work can all be relatively well fit by single \sersic profiles, we note that the inside-out growth pattern we derive in this work is totally model independent. 
The fact that the \sersic profiles can fit all the BCG profiles relatively well only demonstrates the large freedom the \sersic model provides in fitting galaxy profiles but does not invalidate the inside-out growth pattern itself.  
However, as we have noted, since we are only probing the ETGs that grow mostly through dry mergers, we cannot rule out the possibility that gas-rich mergers could also produce an inside-out growth as observed here. 
In this sense, we do concur with \citet{Saracco12} that the inside-out growth pattern reported here does not automatically provide a direct link to the past merger history of the ETGs.
Another point that is worth noting is that the inside-out growth only describes the growth of the mass profile of massive ETGs, and it does not necessarily mean the actual stellar content of the central region of massive ETGs remains unchanged while the galaxy grows bigger.

\section{SUMMARY AND CONCLUSIONS}
Using the ACS observations of 37 rich clusters at $0.3<z<0.9$, we have measured the surface brightness profiles of their BCGs and investigated how the shape of the profiles changes with the mass of the BCGs and how they evolve with redshift.
\begin{enumerate}
\item {\it In the majority of the cases, the profiles of the BCGs can be well fitted by a single \sersic model with a median \sersic index, $n\sim6$, and half-light radius, $r_{e}\sim30$ kpc.} 
In some cases, two-component models (2deV and deV+Exp), do give slightly better fits to the extended outer envelopes. 
However, all the model-fitting parameters are very susceptible to the uncertainties in the sky background measurements and contamination from ICL.
Also, the half-light radius, $r_{e}$, and \sersic index, $n$, given by the best-fit \sersic model are coupled. 
These issues make direct comparisons of the individual model parameters between different studies, sometimes even within the same study, very susceptible to systematics in the profile fitting and great care should be taken in these comparisons.

\item {\it For massive ETGs and BCGs, direct comparisons of the central surface brightnesses and the outer profile slopes are more robust than comparisons of the model-fitting parameters. 
When the mass of these galaxies increases, the central surface brightness only increases slowly while the outer profile slope becomes much shallower.} 
For the BCGs, their central surface brightnesses within 1 kpc span a narrow range, and the corresponding central stellar mass densities increase with the mass of galaxies as $\rho_{\rm 1kpc}\propto M_{*}^{0.25}$.
The average central mass density of the BCGs, $\sim10^{10.2}M_{\sun}~ \mbox{kpc}^{-3}$, agrees very well with the measurement from the local massive ETGs and higher-redshift ones ($1<z<2$) \citep{Saracco12,Tiret11}, suggesting little evolution in the central mass densities. 
The slopes of the outer profile become shallower with the increase of mass as $\alpha_{r^{1/4}} \propto M_{*}^{-2.5}$ and $\alpha_{\log r} \propto M_{*}^{-1.5}$, and the trend continues to the less massive ETGs, without apparent distinction between BCGs and non-BCGs.
These results strongly support an inside-out growth for massive ETGs, which is likely to be driven by dry mergers. 
However, limited by our galaxy sample, we cannot conclude that such a growth pattern is truly unique to dry mergers.

\item {\it The BCGs have a Kormendy relation of $\langle \mu \rangle = 18.01 + 3.50 \log r_{e}$, and its slope and normalization both agree very well with the Kormendy relation of the local BCGs ($z \lesssim 0.1$), except for having much smaller $r_{e}$ than local BCGs.}
We argue that such consistency, despite systematics in the profile fitting of different studies, is due to the facts that the central surface brightnesses of the massive ETGs span only a small range and the slopes of their outer profiles are not very different, causing the Kormendy relation to trace the similar outer profiles of these galaxies.
The slope of the Kormendy relation, in this case, merely reflects the average slope of the profiles of these galaxies.
While the systematic error in the sky background measurement affects both $r_{e}$ and $n$ estimates in profile fitting, it does not change the slope of the profile too much and therefore does not affect the Kormendy relation significantly. 
However, this also means that the Kormendy relation becomes insensitive to the subtle evolution in the BCG profiles.

\item {\it We find that the stellar mass of the BCG correlates strongly with the richness of the clusters.
Furthermore, from $z=0.5$ to $z=0$, the mass of BCGs has increased by at least a factor of 1.5, consistent with the evolution predicted by the $\Lambda$CDM model.}
The stellar mass of the BCGs increases with the cluster mass as $\propto M^{0.6\pm0.1}_{\rm cluster}$.
When we compare BCGs at $z\sim0.5$ with the local BCGs that are in clusters about two times more massive, the cluster mass growth rate expected for this redshift range, we find the outer profile of the BCGs gets shallower, and the isophotal mass increases by 1.5.
This lower limit on the mass growth rate of the BCGs agrees very well with the factor of two growth predicted by the $\Lambda$CDM model. 
Our study shows that the bias in the cluster sample, together with the fact that the model-dependent profile parameters are poor indicators of profile evolution, are probably the main reasons why some of the previous studies failed to find evolution in BCG properties \citep[e.g.,][]{Collins09,Stott10}. 

\end{enumerate}
  
\acknowledgments
We thank the referee for their constructive comments.
L.B. would like to thank John Dubinski and Suresh Sivanandam for helpful discussions.
H.Y.'s research is supported by grants from the Natural Science and Engineering Research Council of Canada and the Canada Research Chair Program.
L.F.B.'s research is supported by Proyecto FONDECYT 1120676.
The paper is based on observations made with the NASA/ESA $Hubble Space Telescope$ and obtained from the Hubble Legacy Archive, which is a collaboration between the Space Telescope Science Institute (STScI/NASA), the Space Telescope European Coordinating Facility (ST-ECF/ESA) and the Canadian Astronomy Data Centre (CADC/NRC/CSA).

\clearpage
\appendix
\section{Profile-Fitting Results of Different Models} 
In this section, we list the best-fit parameters of all four models for the BCG surface brightness profiles given by GALFIT.
The results of the single \sersic model are listed in Table A.1, and the results of the single \dev model are in Table A.2.
The two-component models, 2deV and deV+Exp, are listed in Tables A.3 and A.4 respectively. 
\begin{deluxetable}{llrrrr}
\tablecolumns{6}
\tablewidth{15pc}
\tablecaption{Single Sersic Parameters}
\tablehead{
\colhead{Cluster} & \colhead{$I814$} & \colhead{$r_{e}$} & \colhead{$n$} & \colhead{$b/a$} & \colhead{$\chi^{2}$}\\
& \colhead{(mag)} & \colhead{(kpc)} & & & 
}
\tablecomments{Column 1: galaxy cluster; Column 2: total magnitude from the best-fitting model in $I814$ band; Column 3: half-light radius in kpc; Column 4: \sersic index; Column 5: minor-to-major axis ratio; Column 6: reduced $\chi^{2}$ of the fit.}
\startdata
\hline
RCS$1102-05$&17.89&  11.2$\pm$0.1& 6.71$\pm$0.03&0.91&1.054\\
RCS$2239-60$&17.64&  26.3$\pm$0.3& 5.16$\pm$0.03&0.76&1.075\\
RCS$0444-28$&17.21&  27.0$\pm$0.2& 5.69$\pm$0.02&0.80&0.974\\
RCS$0351-09$&17.24&   6.7$\pm$0.0& 4.54$\pm$0.01&0.95&1.025\\
RCS$0518-43$&17.75&  28.8$\pm$0.6&10.15$\pm$0.07&0.84&1.078\\
RCS$1102-03$&16.48&  48.9$\pm$0.5& 7.67$\pm$0.03&0.84&0.996\\
RCS$0224-02$&18.07&   6.9$\pm$0.1& 5.53$\pm$0.03&0.91&1.061\\
RCS$0515-43$&17.57&  25.8$\pm$0.3& 7.22$\pm$0.04&0.76&1.035\\
RCS$0928+36$&17.69&  15.8$\pm$0.1& 4.34$\pm$0.02&0.87&1.066\\
RCS$1452+08$&17.28&  21.6$\pm$0.1& 4.37$\pm$0.02&0.90&0.998\\
RCS$1319-02$&18.89&   8.1$\pm$0.2&10.15$\pm$0.12&0.84&1.114\\
RCS$1323+30$&17.17&  35.9$\pm$0.3& 5.48$\pm$0.02&0.88&1.009\\
RCS$0511-42$&17.62&  20.6$\pm$0.2& 4.81$\pm$0.02&0.94&1.028\\
RCS$0518-43$&18.07&  37.9$\pm$1.0& 9.10$\pm$0.08&0.83&1.055\\
RCS$1107-05$&18.90&  12.0$\pm$0.2& 5.58$\pm$0.05&0.79&1.134\\
RCS$0519-42$&19.94&   3.5$\pm$0.0& 4.06$\pm$0.04&0.79&1.122\\
RCS$2316-00$&18.89&   9.7$\pm$0.2& 6.14$\pm$0.05&0.91&1.162\\
RCS$0350-08$&18.34&  21.9$\pm$0.4& 7.22$\pm$0.05&0.70&1.071\\
RCS$1108-04$&19.23&  12.3$\pm$0.2& 3.41$\pm$0.03&0.48&1.190\\
RCS$1446+08$&18.41&  92.8$\pm$5.3& 8.78$\pm$0.14&0.69&1.119\\
RCS$1419+53$&18.08&  50.0$\pm$1.1& 6.00$\pm$0.05&0.94&1.054\\
RCS$1104-04$&18.74&  19.4$\pm$0.3& 3.92$\pm$0.03&0.78&1.109\\
RCS$2342-35$&19.15&  52.8$\pm$4.2& 8.94$\pm$0.20&0.88&1.102\\
RCS$1107-05$&19.29&  28.5$\pm$0.9& 4.83$\pm$0.07&0.66&1.118\\
RCS$2152-06$&17.90& 112.9$\pm$5.1&10.22$\pm$0.12&0.87&1.088\\
RCS$1450+08$&18.80&  35.2$\pm$0.9& 5.74$\pm$0.06&0.63&1.161\\
RCS$1122+24$&19.50&  14.6$\pm$0.3& 4.07$\pm$0.05&0.72&1.103\\
RCS$1620+29$&20.52&  11.9$\pm$0.8& 6.56$\pm$0.19&0.80&1.165\\
RCS$0519-44$&19.88&   8.7$\pm$0.1& 3.52$\pm$0.04&0.84&1.158\\
RXJ$0110+19$&16.80&  22.8$\pm$0.1& 4.60$\pm$0.01&0.86&0.946\\
RXJ$0841+64$&15.95& 162.3$\pm$1.8& 5.82$\pm$0.02&0.62&0.926\\
RXJ$1540+14$&17.50&  58.6$\pm$1.1& 8.35$\pm$0.05&0.93&0.974\\
RXJ$0926+12$&17.98&  27.3$\pm$0.3& 4.37$\pm$0.02&0.63&1.070\\
RXJ$2328+14$&17.65&  44.4$\pm$0.8& 7.26$\pm$0.04&0.96&1.036\\
RXJ$0056-27$&18.14&  29.5$\pm$0.3& 4.37$\pm$0.02&0.81&1.013\\
RXJ$0847+34$&17.85&  41.4$\pm$0.6& 5.75$\pm$0.03&0.90&1.006\\
RXJ$1354-02$&17.26&  38.6$\pm$0.4& 5.43$\pm$0.02&0.79&0.995\\
\enddata
\label{tab_sers}
\end{deluxetable}

\begin{deluxetable}{lrrrr}
\tablecolumns{5}
\tablewidth{15pc}
\tablecaption{\dev Parameters}
\tablehead{
\colhead{Cluster} & \colhead{$I814$} & \colhead{$r_{e}$}  & \colhead{$b/a$} & \colhead{$\chi^{2}$}\\
& \colhead{(mag)} & \colhead{(kpc)} & & 
}
\tablecomments{Column 1: galaxy cluster; Column 2: total magnitude from the best-fitting model in $I814$ band; Column 3: half-light radius in kpc; Column 4: the minor-to-major axis ratio; Column 5: reduced $\chi^{2}$ of the fit.}
\startdata
\hline
RCS$1102-05$&18.28&   5.2$\pm$0.0&0.92&1.072\\
RCS$2239-60$&17.86&  17.6$\pm$0.1&0.77&1.079\\
RCS$0444-28$&17.49&  15.6$\pm$0.0&0.80&0.992\\
RCS$0351-09$&17.30&   5.9$\pm$0.0&0.95&1.031\\
RCS$0518-43$&18.54&   5.3$\pm$0.0&0.86&1.130\\
RCS$1102-03$&17.11&  14.3$\pm$0.0&0.85&1.075\\
RCS$0224-02$&18.26&   4.8$\pm$0.0&0.91&1.071\\
RCS$0515-43$&18.07&   9.7$\pm$0.0&0.77&1.068\\
RCS$0928+36$&17.76&  13.9$\pm$0.0&0.87&1.067\\
RCS$1452+08$&17.35&  18.8$\pm$0.1&0.90&0.999\\
RCS$1319-02$&19.45&   2.3$\pm$0.0&0.84&1.136\\
RCS$1323+30$&17.47&  21.0$\pm$0.1&0.89&1.024\\
RCS$0511-42$&17.78&  15.4$\pm$0.0&0.95&1.033\\
RCS$0518-43$&18.74&   9.2$\pm$0.0&0.84&1.082\\
RCS$1107-05$&19.14&   7.6$\pm$0.0&0.79&1.138\\
RCS$0519-42$&19.85&   3.5$\pm$0.0&0.79&1.122\\
RCS$2316-00$&19.15&   5.7$\pm$0.0&0.91&1.168\\
RCS$0350-08$&18.80&   8.6$\pm$0.0&0.72&1.084\\
RCS$1108-04$&19.13&  14.6$\pm$0.1&0.48&1.191\\
RCS$1446+08$&19.30&  16.3$\pm$0.1&0.72&1.127\\
RCS$1419+53$&18.52&  22.9$\pm$0.1&0.94&1.059\\
RCS$1104-04$&18.72&  20.0$\pm$0.1&0.78&1.109\\
RCS$2342-35$&20.07&   8.8$\pm$0.1&0.91&1.107\\
RCS$1107-05$&19.48&  20.3$\pm$0.2&0.67&1.119\\
RCS$2152-06$&18.95&  14.0$\pm$0.1&0.86&1.110\\
RCS$1450+08$&19.14&  18.8$\pm$0.1&0.64&1.163\\
RCS$1122+24$&19.51&  14.3$\pm$0.1&0.73&1.103\\
RCS$1620+29$&20.95&   5.1$\pm$0.1&0.81&1.166\\
RCS$0519-44$&19.80&  10.1$\pm$0.1&0.84&1.158\\
RXJ$0110+19$&16.92&  18.6$\pm$0.0&0.86&0.957\\
RXJ$0841+64$&16.40&  74.7$\pm$0.1&0.63&0.948\\
RXJ$1540+14$&18.28&  12.8$\pm$0.0&0.95&1.038\\
RXJ$0926+12$&18.06&  23.7$\pm$0.1&0.62&1.070\\
RXJ$2328+14$&18.30&  13.1$\pm$0.0&0.98&1.063\\
RXJ$0056-27$&18.22&  25.7$\pm$0.1&0.81&1.014\\
RXJ$0847+34$&18.28&  19.5$\pm$0.1&0.90&1.020\\
RXJ$1354-02$&17.57&  22.1$\pm$0.1&0.80&1.009\\
\enddata
\label{tab_dev}
\end{deluxetable}

\begin{deluxetable}{lrrrrrrrrrr}
\tabletypesize{\scriptsize}
\tablewidth{43pc}
\tablecolumns{11}
\tablecaption{Two \dev Parameters}
\tablehead{
\colhead{Cluster} & \colhead{$I814_{\rm total}$} & \colhead{$I814^{1}$} & \colhead{$r_{e}^{1}$}  & \colhead{$b/a^{1}$} & \colhead{P.A.$^{1}$} & \colhead{$I814^{2}$} & \colhead{$r_{e}^{2}$}  & \colhead{$b/a^{2}$} & \colhead{P.A.$^{2}$} & \colhead{$\chi^{2}$}\\
& \colhead{(mag)} & \colhead{(mag)} & \colhead{(kpc)} & & \colhead{(deg)} & \colhead{(mag)} & \colhead{(kpc)} & & \colhead{(deg)} & 
} 
\tablecomments{Column 1: galaxy cluster; Column 2: total magnitude from two \dev models; Columns 3-6: magnitude, half-light radius, minor-to-major axis ratio and position angle of the first \dev component; Columns 7-10: same quantities for the second \dev component; Column 11: reduced $\chi^{2}$ of the fit.}
\startdata
\hline
RCS$1102-05$&17.77&18.75&   3.1$\pm$0.0&0.94& 11&18.33&  56.7$\pm$2.6&0.57& 76&1.046\\
RCS$2239-60$&17.58&19.09&   7.2$\pm$0.1&0.86&  7&17.89&  53.9$\pm$2.3&0.49& 85&1.069\\
RCS$0444-28$&17.19&18.75&   5.7$\pm$0.1&0.84& 70&17.48&  48.5$\pm$1.3&0.64& 19&0.967\\
RCS$0351-09$&16.11&17.38&   5.4$\pm$0.0&0.93& 59&16.51& 888.7$\pm$85.3&0.44& 30&1.010\\
RCS$0518-43$&17.89&19.61&   1.6$\pm$0.0&0.81& 73&18.14&  31.0$\pm$0.5&0.86& 84&1.072\\
RCS$1102-03$&16.23&17.86&   6.6$\pm$0.0&0.83& 37&16.50& 146.2$\pm$3.0&0.81& 26&0.984\\
RCS$0224-02$&18.03&19.10&   2.5$\pm$0.1&0.77& 27&18.54&  15.4$\pm$0.7&0.86& 77&1.057\\
RCS$0515-43$&17.52&19.02&   3.6$\pm$0.0&0.82& 14&17.83&  54.3$\pm$1.5&0.67&  7&1.031\\
RCS$0928+36$&17.62&18.93&   6.8$\pm$0.1&0.90&  4&18.00&  29.1$\pm$0.5&0.75& 48&1.055\\
RCS$1452+08$&17.07&17.97&  11.9$\pm$0.1&0.90& 77&17.70&  92.7$\pm$2.8&0.51&  4&0.987\\
RCS$1319-02$&18.73&19.94&   1.2$\pm$0.0&0.79& 64&19.16&  33.3$\pm$1.3&0.85& 30&1.110\\
RCS$1323+30$&17.16&18.98&   7.3$\pm$0.1&0.83& 29&17.39&  55.4$\pm$1.3&0.72& 52&1.002\\
RCS$0511-42$&17.43&18.61&   8.9$\pm$0.1&0.77& 76&17.87&  75.1$\pm$2.0&0.52&  8&1.005\\
RCS$0518-43$&18.42&20.88&   1.0$\pm$0.0&0.78& 34&18.54&  21.7$\pm$0.3&0.78& 24&1.052\\
RCS$1107-05$&18.77&19.75&   4.2$\pm$0.1&0.86& 21&19.34&  44.1$\pm$4.0&0.62& 24&1.133\\
RCS$0519-42$&19.89&20.31&   2.8$\pm$0.1&0.77& 65&21.12&   8.6$\pm$1.5&0.80& 78&1.122\\
RCS$2316-00$&18.96&21.27&   1.2$\pm$0.0&0.49& 63&19.10&  10.1$\pm$0.2&0.95& 21&1.160\\
RCS$0350-08$&18.13&19.41&   4.4$\pm$0.1&0.74& 22&18.53&  87.0$\pm$3.8&0.60& 18&1.068\\
RCS$1108-04$&19.08&19.25&  14.7$\pm$0.2&0.43& 83&21.19&  28.5$\pm$6.9&0.58&  1&1.190\\
RCS$1446+08$&18.75&21.34&   2.3$\pm$0.1&0.80&  9&18.86&  56.5$\pm$1.8&0.52& 64&1.116\\
RCS$1419+53$&18.20&20.78&   4.6$\pm$0.2&0.86& 10&18.31&  45.8$\pm$1.4&0.97& 26&1.053\\
RCS$1104-04$&18.61&19.90&   9.9$\pm$0.3&0.93& 71&19.01&  40.6$\pm$1.1&0.53& 16&1.106\\
RCS$2342-35$&19.30&21.35&   2.6$\pm$0.1&0.86& 67&19.48&  56.1$\pm$4.4&0.65&  2&1.101\\
RCS$1107-05$&19.21&20.87&   8.1$\pm$0.7&0.72& 87&19.48&  52.1$\pm$5.6&0.50& 53&1.118\\
RCS$2152-06$&18.18&20.41&   3.5$\pm$0.0&0.63& 47&18.33&  81.4$\pm$2.0&0.60& 27&1.079\\
RCS$1450+08$&18.76&20.63&   4.7$\pm$0.1&0.92& 56&18.97&  62.9$\pm$2.7&0.44& 41&1.157\\
RCS$1122+24$&19.39&20.41&   8.1$\pm$0.4&0.80& 58&19.92&  33.0$\pm$2.6&0.55& 22&1.102\\
RCS$1620+29$&20.17&21.53&   2.8$\pm$0.1&0.80& 85&20.53&  52.0$\pm$7.6&0.85& 71&1.164\\
RCS$0519-44$&19.82&20.74&   9.7$\pm$1.0&0.81& 57&20.42&  10.1$\pm$0.8&0.81& 73&1.157\\
RXJ$0110+19$&16.69&17.50&  11.5$\pm$0.1&0.94& 10&17.38&  81.0$\pm$2.4&0.56& 37&0.925\\
RXJ$0841+64$&16.12&19.23&   8.6$\pm$0.1&0.92& 83&16.18& 137.6$\pm$1.0&0.55& 47&0.916\\
RXJ$1540+14$&17.46&19.31&   4.5$\pm$0.0&0.96& 90&17.68&  93.2$\pm$2.1&0.87& 54&0.962\\
RXJ$0926+12$&17.92&19.63&   9.1$\pm$0.2&0.74& 49&18.17&  43.3$\pm$0.9&0.53& 71&1.068\\
RXJ$2328+14$&17.56&19.35&   5.3$\pm$0.1&0.86& 73&17.79&  89.1$\pm$2.4&0.72& 23&1.029\\
RXJ$0056-27$&18.14&22.91&   1.1$\pm$0.1&0.44&  4&18.15&  30.4$\pm$0.3&0.78& 80&1.009\\
RXJ$0847+34$&15.95&18.66&  13.7$\pm$0.1&0.87& 35&16.04&1416.1$\pm$122.3&0.51& 34&0.987\\
RXJ$1354-02$&17.34&19.80&   5.5$\pm$0.2&0.68& 42&17.46&  39.4$\pm$0.8&0.79&  8&0.993\\
\enddata
\label{tab_2dev}
\end{deluxetable}

\begin{deluxetable}{lrrrrrrrrrr}
\tabletypesize{\scriptsize}
\tablewidth{43pc}
\tablecolumns{11}
\tablecaption{\dev + Exponential Parameters}
\tablehead{
\colhead{Cluster} & \colhead{$I814_{\rm total}$} & \colhead{$I814^{1}$} & \colhead{$r_{e}^{1}$}  & \colhead{$b/a^{1}$} & \colhead{P.A.$^{1}$} & \colhead{$I814^{2}$} & \colhead{$r_{e}^{2}$}  & \colhead{$b/a^{2}$} & \colhead{P.A.$^{2}$} & \colhead{$\chi^{2}$}\\
& \colhead{(mag)} & \colhead{(mag)} & \colhead{(kpc)} & & \colhead{(deg)} & \colhead{(mag)} & \colhead{(kpc)} & & \colhead{(deg)} & 
}
\tablecomments{Column 1: galaxy cluster; Column 2: total magnitude from deV+Exp components; Columns 3-6: magnitude, half-light radius, minor-to-major axis ratio and position angle of the \dev component; Columns 7-10: same quantities for the Exponential component; Column 11: reduced $\chi^{2}$ of the fit.}

\startdata
\hline
RCS$1102-05$&17.98&18.51&   3.8$\pm$0.0&0.95& 54&19.02&  42.5$\pm$0.7&0.43& 73&1.043\\
RCS$2239-60$&17.86&18.41&   9.7$\pm$0.1&0.88& 81&18.86&  41.8$\pm$0.6&0.37& 84&1.065\\
RCS$0444-28$&17.42&17.90&   9.6$\pm$0.0&0.86& 35&18.53&  49.2$\pm$0.6&0.49& 20&0.969\\
RCS$0351-09$&16.98&17.36&   5.5$\pm$0.0&0.95& 60&18.29& 119.5$\pm$4.1&0.49& 25&1.011\\
RCS$0518-43$&18.15&19.01&   2.7$\pm$0.0&0.85& 73&18.81&  25.3$\pm$0.2&0.78& 88&1.072\\
RCS$1102-03$&16.70&17.49&   8.9$\pm$0.0&0.84& 34&17.41&  76.3$\pm$0.6&0.70& 34&0.985\\
RCS$0224-02$&18.17&18.49&   3.6$\pm$0.0&0.88& 32&19.67&  19.8$\pm$0.3&0.72& 72&1.058\\
RCS$0515-43$&17.79&18.47&   5.8$\pm$0.0&0.78& 10&18.62&  41.5$\pm$0.5&0.65&  9&1.031\\
RCS$0928+36$&17.54&17.92&  11.5$\pm$0.0&0.88& 35&18.87&  88.6$\pm$2.6&0.43& 62&1.056\\
RCS$1452+08$&17.27&17.61&  14.2$\pm$0.1&0.97& 22&18.70&  77.8$\pm$2.1&0.29&  2&0.992\\
RCS$1319-02$&18.98&19.66&   1.6$\pm$0.0&0.82& 59&19.81&  23.5$\pm$0.4&0.84& 36&1.110\\
RCS$1323+30$&17.44&18.01&  11.6$\pm$0.1&0.95& 42&18.40&  46.9$\pm$0.4&0.60& 52&0.998\\
RCS$0511-42$&17.61&18.10&  10.9$\pm$0.0&0.94& 63&18.71&  75.4$\pm$1.1&0.36&  7&1.007\\
RCS$0518-43$&18.60&19.39&   3.9$\pm$0.0&0.86&  4&19.32&  27.6$\pm$0.2&0.65& 32&1.056\\
RCS$1107-05$&18.95&19.36&   5.7$\pm$0.0&0.81& 23&20.21&  42.2$\pm$1.4&0.51& 18&1.133\\
RCS$0519-42$&19.88&19.93&   3.4$\pm$0.0&0.76& 73&23.19&   3.7$\pm$0.3&0.59&  9&1.122\\
RCS$2316-00$&19.06&19.56&   3.6$\pm$0.0&0.84& 59&20.15&  18.7$\pm$0.4&0.71& 29&1.161\\
RCS$0350-08$&18.39&19.06&   6.0$\pm$0.0&0.72& 21&19.24&  65.9$\pm$1.1&0.49& 19&1.069\\
RCS$1108-04$&19.16&19.25&  14.1$\pm$0.1&0.56& 83&21.92&   8.8$\pm$0.3&0.18& 83&1.189\\
RCS$1446+08$&19.20&20.37&   4.5$\pm$0.1&0.85& 53&19.65&  34.3$\pm$0.5&0.44& 65&1.115\\
RCS$1419+53$&18.48&19.12&  12.8$\pm$0.1&0.92&  3&19.35&  42.6$\pm$0.8&0.89& 68&1.053\\
RCS$1104-04$&18.55&18.63&  35.7$\pm$0.6&0.65& 15&21.44&   2.8$\pm$0.0&0.69& 70&1.105\\
RCS$2342-35$&19.68&20.66&   4.4$\pm$0.1&0.94& 32&20.25&  37.0$\pm$1.4&0.55&  6&1.101\\
RCS$1107-05$&19.28&19.85&  13.8$\pm$0.2&0.70& 67&20.26&  77.4$\pm$3.8&0.28& 54&1.117\\
RCS$2152-06$&18.65&19.77&   5.2$\pm$0.0&0.80& 36&19.13&  45.4$\pm$0.5&0.49& 33&1.077\\
RCS$1450+08$&19.02&19.65&  10.4$\pm$0.1&0.70& 42&19.90&  57.4$\pm$1.5&0.35& 40&1.158\\
RCS$1122+24$&19.40&19.65&  12.4$\pm$0.2&0.73& 40&21.14&  79.3$\pm$8.7&0.18&  8&1.102\\
RCS$1620+29$&20.28&21.15&   4.0$\pm$0.1&0.79& 83&20.93&  49.5$\pm$3.2&0.68& 89&1.164\\
RCS$0519-44$&19.78&19.85&  12.0$\pm$0.2&0.79& 66&22.87&   2.5$\pm$0.1&0.77& 18&1.157\\
RXJ$0110+19$&16.74&16.81&  30.4$\pm$0.1&0.82& 34&19.68&   2.4$\pm$0.0&0.92& 49&0.898\\
RXJ$0841+64$&16.39&16.93&  43.8$\pm$0.1&0.68& 47&17.40& 161.2$\pm$1.3&0.31& 51&0.926\\
RXJ$1540+14$&17.86&18.74&   7.2$\pm$0.0&0.95& 82&18.50&  60.3$\pm$0.5&0.75& 52&0.963\\
RXJ$0926+12$&17.94&17.98&  33.0$\pm$0.3&0.59& 66&21.63&   1.8$\pm$0.0&0.88& 12&1.068\\
RXJ$2328+14$&17.97&18.81&   7.4$\pm$0.0&0.96& 89&18.64&  51.9$\pm$0.5&0.63& 20&1.029\\
RXJ$0056-27$&18.37&18.86&  14.1$\pm$0.1&0.90& 69&19.46&  33.3$\pm$0.4&0.52& 85&1.007\\
RXJ$0847+34$&17.75&17.82&  53.6$\pm$0.5&0.87&  1&20.78&   2.2$\pm$0.0&0.79& 54&0.979\\
RXJ$1354-02$&17.39&17.42&  31.0$\pm$0.2&0.80& 16&21.37&   1.5$\pm$0.0&0.62& 33&0.991\\
\enddata
\label{tab_n4n1}
\end{deluxetable}

\section{The 1D Surface Brightness Profiles for the BCG and best-fit Models} 
\begin{figure}
\epsscale{1.2}
\plotone{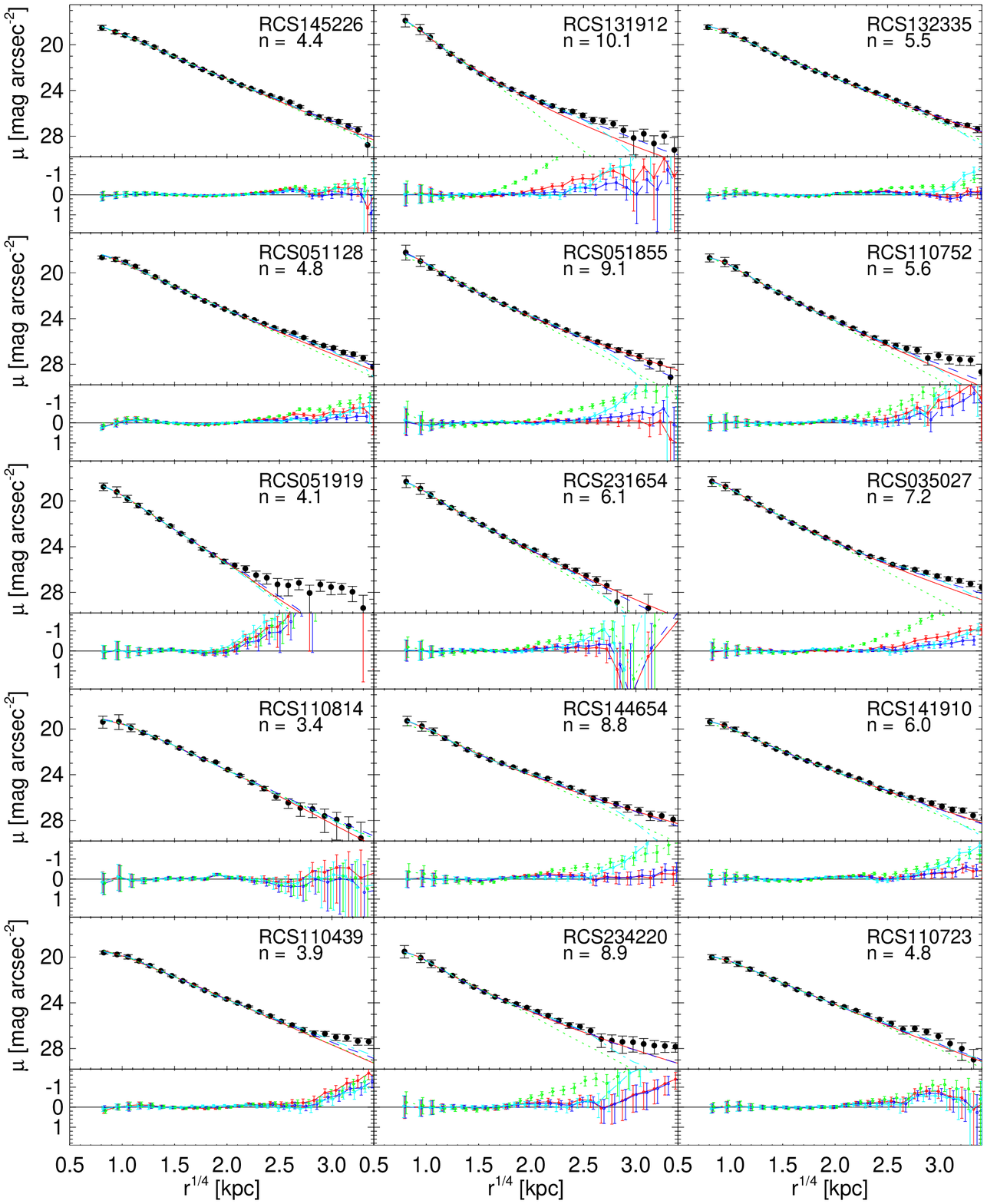}
\caption{1D surface brightness profiles of BCGs and the best-fit models.
Black filled circles are data.
The red solid curve is the single \sersic model, the blue dashed curve is the 2deV model, the cyan dash-dotted curve is the deV+Exp model, and the green dotted curve is the single deV model. 
In the bottom panels, the residuals of each of the model fittings are shown.
}
\label{fig_profile_adx}
\end{figure}

As a continuation to Figure~\ref{fig_profile}, the 1D surface brightness profiles of the rest of the 35 BCGs in our sample are shown in Figure~\ref{fig_profile_adx}. 

\begin{figure}
\epsscale{1.2}
\plotone{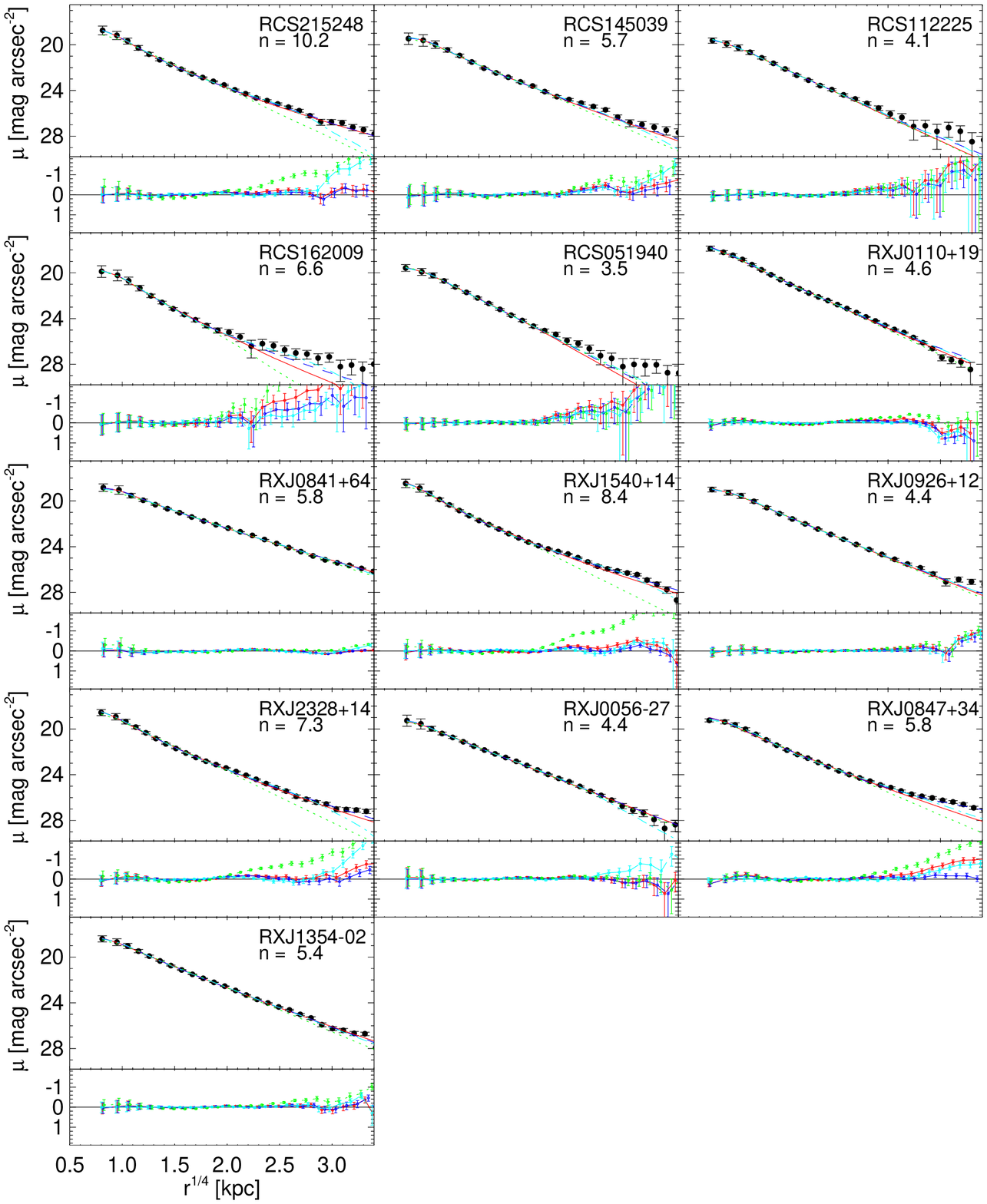}
\renewcommand\thefigure{B1-CONTINUED}
\caption{1D surface brightness profiles of BCGs and the best-fit models.
Black filled circles are data.
The red solid curve is the single \sersic model, the blue dashed curve is the 2deV model, the cyan dash-dotted curve is the deV+Exp model and the green dotted curve is the single deV model.
In the bottom panels, the residuals of each of the model fittings are shown.
}
\end{figure}

\end{document}